\documentclass[electronic]{vgtc}             

\graphicspath{{figures/}{pictures/}{images/}{./}} 
\usepackage{pifont}
\newcommand{\cmark}{\ding{51}}%
\newcommand{\xmark}{\ding{55}}%
\usepackage{times}                     
\usepackage{tabu}                      
\usepackage{booktabs}                  
\usepackage{lipsum} 
\usepackage{mwe}        
\usepackage{subcaption}
\usepackage{multirow}
\usepackage{array}
\usepackage{lscape}
\usepackage{subcaption}  
\usepackage{amsmath}
\usepackage{amssymb}

\usepackage{mathptmx}   
\usepackage{newtxtext,newtxmath}

\onlineid{1836}

\vgtccategory{Research}

\vgtcinsertpkg

\newcolumntype{P}[1]{>{\centering\arraybackslash}p{#1}}

\title{``Just stop doing everything for now!'': Understanding security attacks in remote collaborative mixed reality}

\author{Maha Sajid\thanks{e-mail: mahas@vt.edu}\\ %
\and Syed Ibrahim Mustafa Shah Bukhari\thanks{e-mail:simsb@vt.edu} %
\and Bo Ji\thanks{e-mail:boji@vt.edu}
\and Brendan David-John\thanks{e-mail:bmdj@vt.edu}}
\affiliation{\scriptsize Virginia Tech}

\teaser{
    \centering
    \begin{subfigure}[b]{0.24\textwidth}
        \centering
        \includegraphics[width = 4cm, height=4cm]{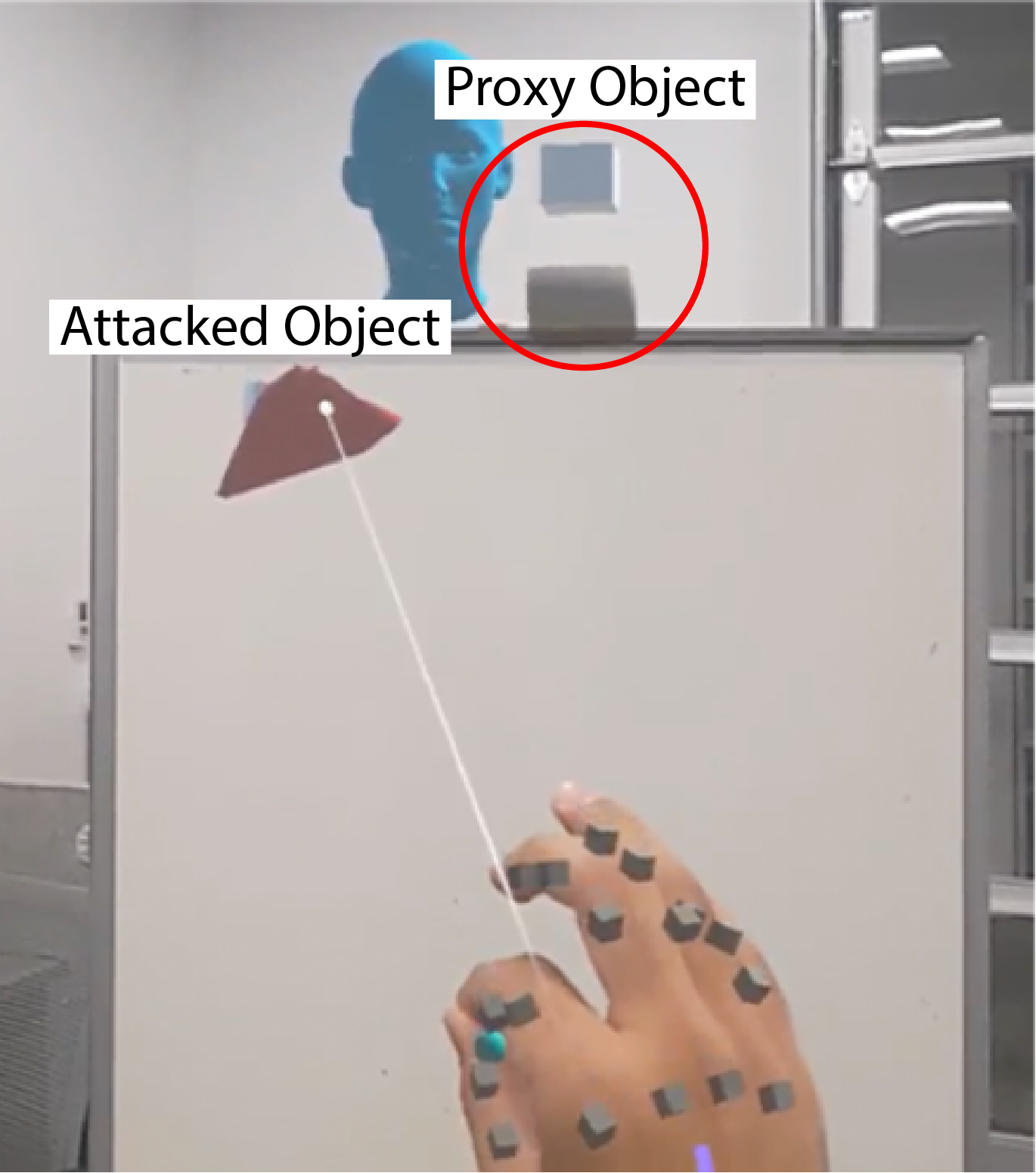}
        \caption{Click Redirection (Before)}
        \label{fig:sub1}
    \end{subfigure}
    \hfill
    \begin{subfigure}[b]{0.24\textwidth}
        \centering
        \includegraphics[width = 4cm, height=4cm]{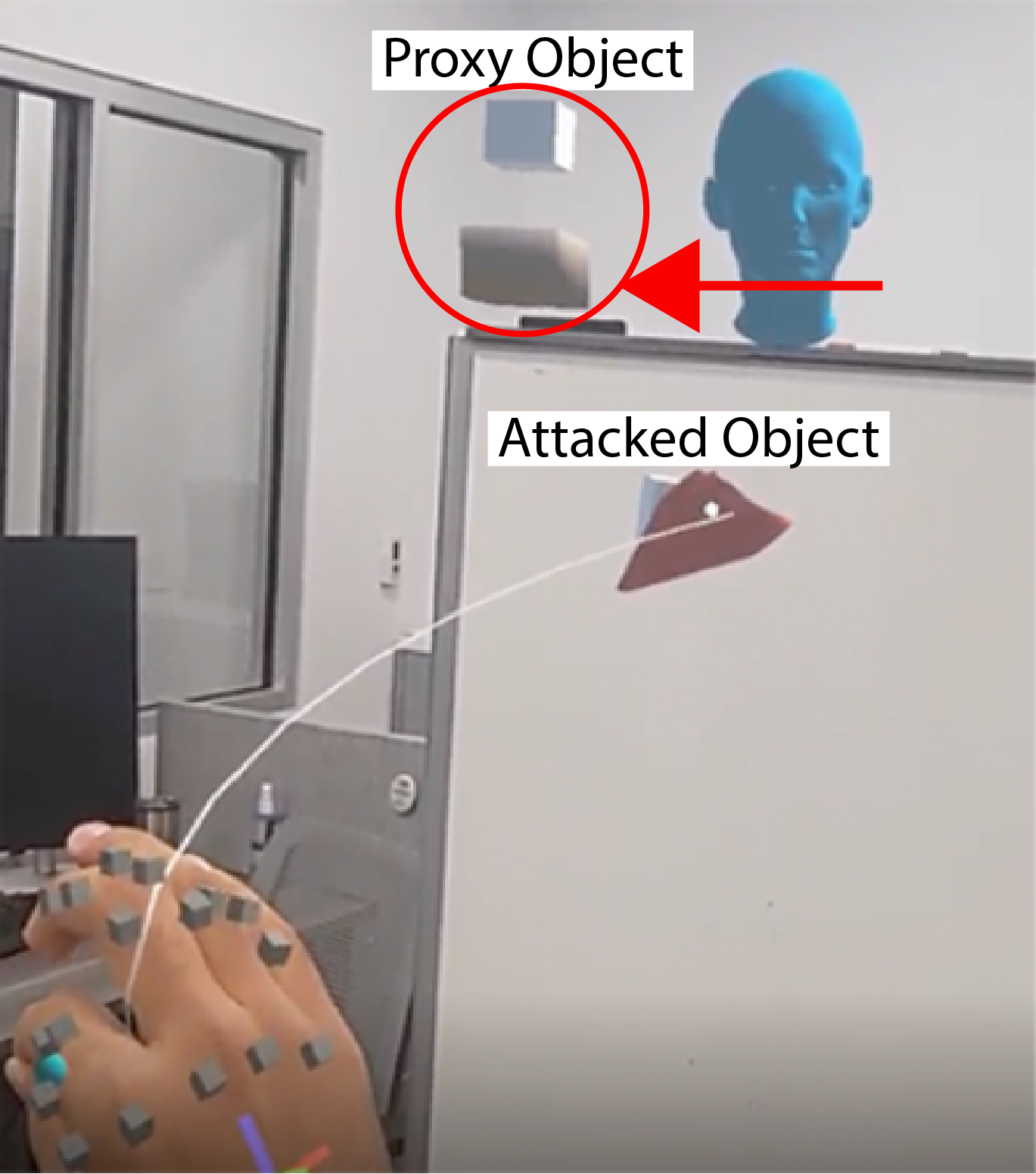}
        \caption{Click Redirection (After)}
        \label{fig:sub2}
    \end{subfigure}
    \hfill
    \begin{subfigure}[b]{0.24\textwidth}
        \centering
        \includegraphics[width = 4cm, height=4cm]{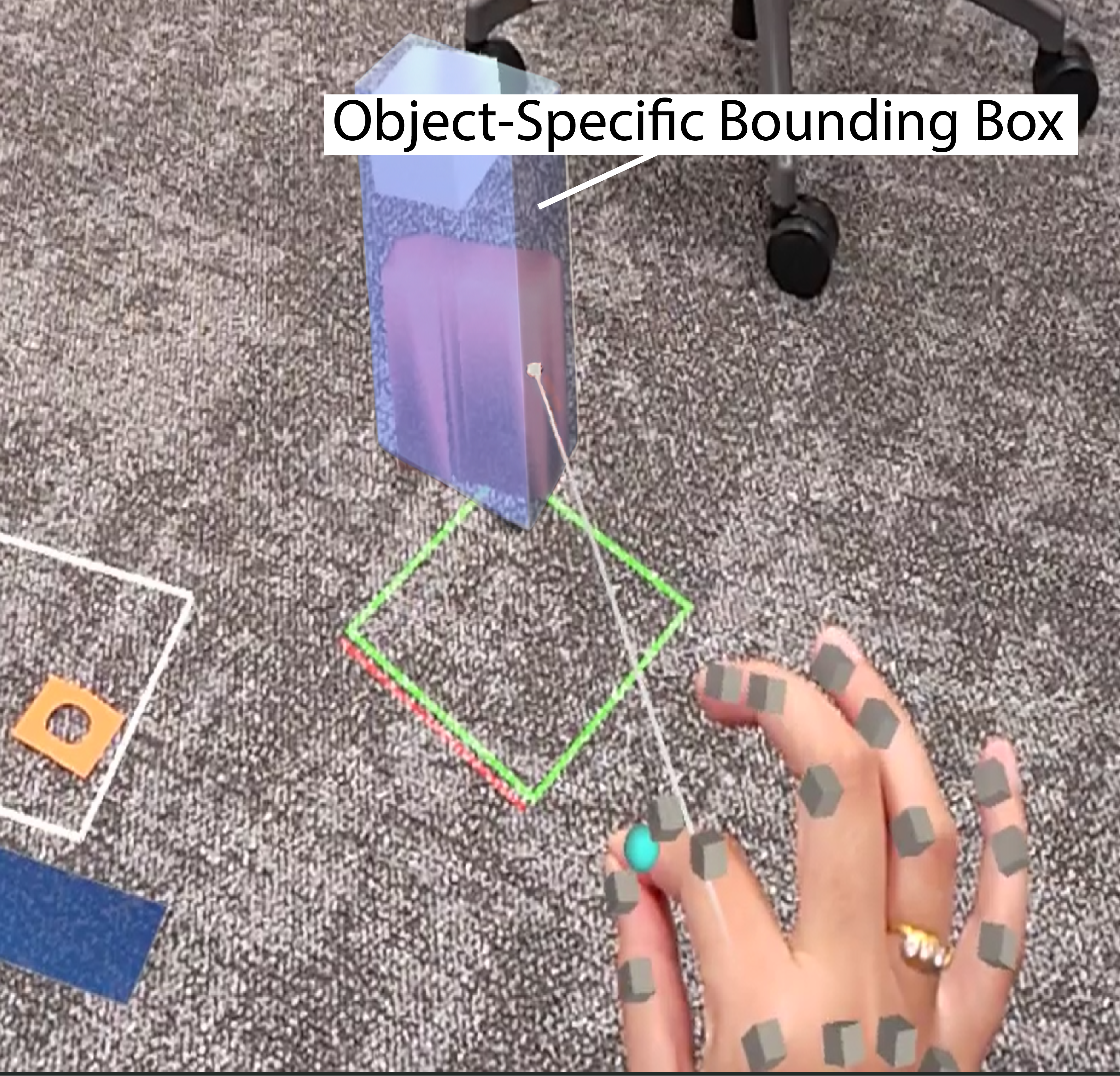}
        \caption{Object Occlusion}
        \label{fig:sub4}
    \end{subfigure}
     \hfill
    \begin{subfigure}[b]{0.24\textwidth}
        \centering
        \includegraphics[width = 4cm, height=4cm]{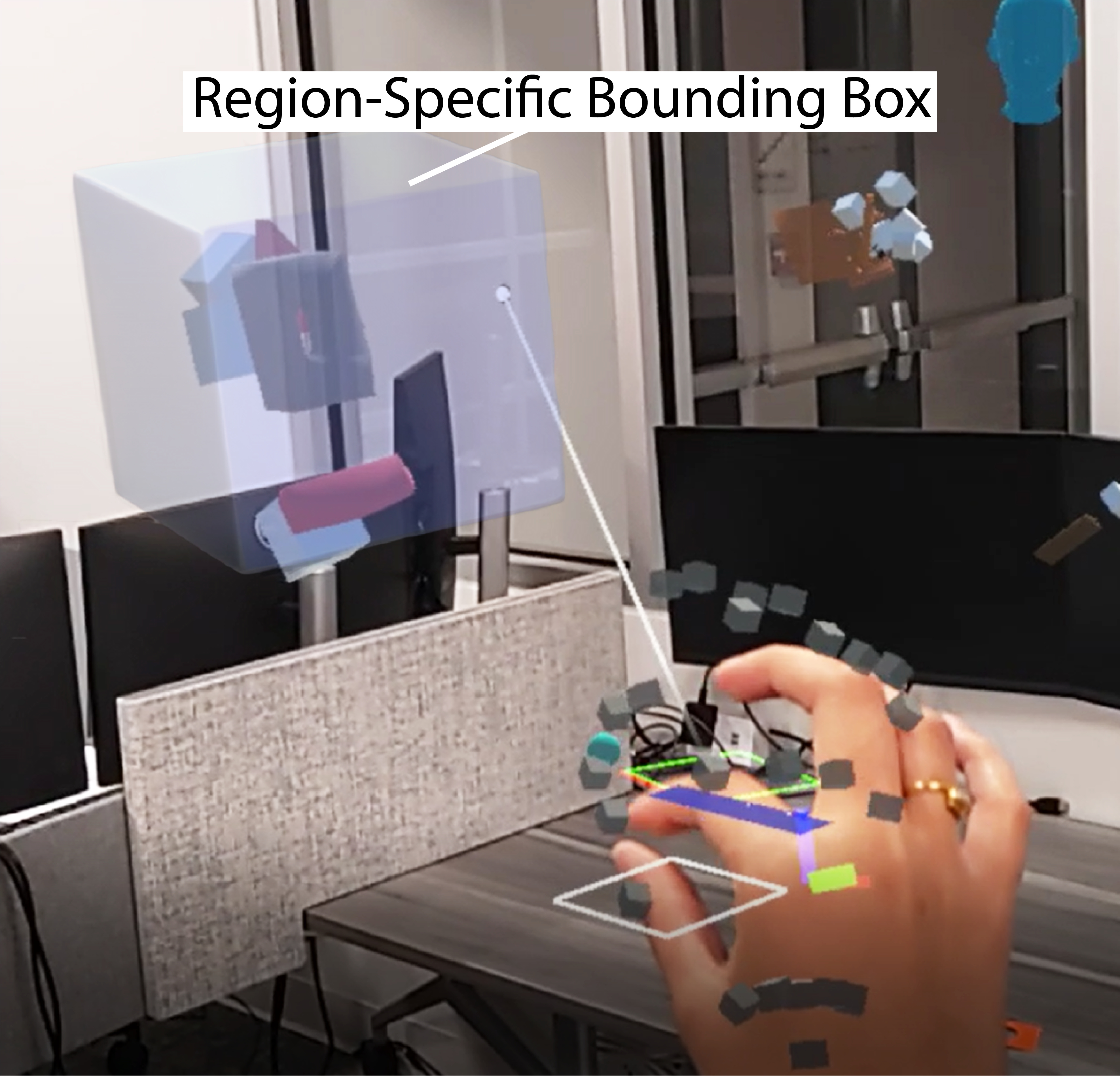}
        \caption{Spatial Occlusion }
        \label{fig:sub3}
    \end{subfigure}
    \caption{Three security attacks in remote collaborative Mixed Reality (MR) considered in this work: click redirection ((a) and (b)), object occlusion (c), and spatial occlusion (d). In (a), the user selects the attacked object and attempts to move it by pinching and dragging; however, (b) shows that a proxy object (encircled) is moved instead (direction indicated by a red arrow). In (c), the user attempts to select an object using a ray cast but cannot interact with it as their ray cast is blocked by a transparent bounding box following the target object (visible here for illustration purposes only). (d) shows spatial occlusion, which is similar to (c), except that the transparent attack bounding box is attached to a fixed region in space instead of a single object.}
    \label{fig:teaser}
}

\abstract{
    Mixed Reality (MR) devices are being increasingly adopted across a wide range of real-world applications, ranging from education and healthcare to remote work and entertainment. However, the unique immersive features of MR devices, such as 3D spatial interactions and the encapsulation of virtual objects by invisible elements, introduce new vulnerabilities leading to interaction obstruction and misdirection. We implemented latency, click redirection, object occlusion, and spatial occlusion attacks within a remote collaborative MR platform using the Microsoft HoloLens 2 and evaluated user behavior and mitigations through a user study. We compared responses to MR-specific attacks, which exploit the unique characteristics of remote collaborative immersive environments, and traditional security attacks implemented in MR. Our findings indicate that users generally exhibit lower recognition rates for immersive attacks (e.g., spatial occlusion) compared to attacks inspired by traditional ones (e.g., click redirection). Our results demonstrate a clear gap in user awareness and responses when collaborating remotely in MR environments. Our findings emphasize the importance of training users to recognize potential threats and enhanced security measures to maintain trust in remote collaborative MR systems.

} 


\begin{document}


\firstsection{Introduction}
\maketitle

Mixed Reality (MR) integrates real and virtual elements to create environments where physical and digital objects can interact in real-time. There has been growing adoption and interest in collaborative MR systems driven by applications in healthcare, education, and engineering ~\cite{10.3934/ElectrEng.2019.2.181,10.1145/3132787.3139200,kaufmann2003collaborative, mourtzis2021collaborative, santos2007improve}. 

Collaborative MR can be either co-located or remote. Remote MR applications are especially intriguing due to their practical applications, complex threat landscape, and greater reliance on technology for collaboration compared to co-located settings~\cite{10.1145/3359626}. Users rely heavily on virtual avatars and object interactions to interpret collaborators’ actions. Remote collaborations increase the potential for undetected security threats compared to co-located settings where physical presence can help mitigate misunderstandings. Additionally, the growing demand for remote work and global collaboration emphasizes the importance of studying remote collaborative MR environments and developing secure applications. 

While the general acceptance of remote collaboration has increased post-COVID-19~\cite{10.3389/feduc.2021.711619}, the 2D nature of traditional computing devices and corresponding remote collaboration tools limit the adoption of collaborative applications, especially in hardware-dependent and high-risk industries such as manufacturing and healthcare. MR is poised to address these limitations by expanding the breadth and impact of remote collaborations. For example, clinical carts used to support telemedicine allow medical experts to provide care when they are not co-located with the patient~\cite{baker2018telemedicine}. However, most of the cart's functionality relies on repurposed video conferencing tools and basic signal monitoring, requiring a local technician and a nurse to facilitate the visit. In contrast, MR technology offers medical professionals a spatially integrated experience with minimal need for additional resources, allowing them to focus solely on patient care. Given these advantages, MR technology has already seen increased usage in telemedicine~\cite{gasques2021artemis,eom2024accuracy}.

New MR collaboration opportunities also introduce novel security risks due to increased connectivity, immersion, extensive use of sensors, and a lack of user understanding ~\cite{10.1145/3359626, 9963687,krauss2024makes, abraham2022implications}. For example, an attack might alter the environment for one user without affecting the view of others or disrupt communication between users at a critical moment. While there have been initial investigations into the security risks and resulting mitigations in MR devices ~\cite{syal2020threats, chenguser, yang2024can, kilger2021detecting}, there is a lack of analysis and understanding of the security risks introduced by networked remote collaborative MR. The work of~\cite{lebeck2018towards} lays the foundation for analyzing multi-user dynamics in MR environments. However, their work does not expose users to security attacks in practice. In contrast, we aim to fill this gap by studying real-time user behavior during security attacks in remote collaborative MR tasks and analyzing how collaboration dynamics influence the perception and mitigation of security threats. We believe that doing so is important to understand the practical risks associated with these systems and how users respond to security threats within such systems. Therefore, we examine user responses to security attacks within remote collaborative MR to understand the \textit{recognition of security attacks} and \textit{mitigation strategies }from the user's perspective.

To the best of our knowledge, our study is the first to explore user responses to security attacks in situ within a remote collaborative MR environment. Our study also identifies common mitigation techniques that users employ against various attacks and examines how these techniques differ from those that they report using.

\textbf{Contributions:}
Our research provides several contributions towards the security and usability of collaborative MR:
\begin{itemize}  \setlength{\itemsep}{0pt}
    \item We are the first to implement security attacks in a remote collaborative MR environment. We evaluate the impact on attack recognition and mitigation strategies through a user study based on a modified prototype on Microsoft HoloLens~2. 
    \item We identify that users are less likely to recognize attacks unique to MR and attribute them to other factors like technological glitches, partner miscommunication, and user errors. 
    \item We identify a disconnect between users' perceived mitigation efforts and their actual responses when faced with attacks in a remote collaborative MR environment.

\end{itemize}

\section{Related Work}

In this section, we briefly discuss relevant literature that explores collaborative MR applications, security threats in such applications, and users' perspectives on these security threats.

{\textbf{Applications of Collaborative MR:}}\label{RW:Applications} Collaborative MR has been increasingly explored for real-life applications~\cite{gasques2017exploring, dell2022designing, rodrigues2018real, johnson2018holocpr, rajaram2023reframe}. In the medical field, Gasques et al.~\cite{gasques2021artemis} proposed using collaborative MR to guide inexperienced surgeons during emergency medical procedures by introducing an MR system called ARTEMIS. They conducted a user study using ARTEMIS to guide novice surgeons through procedures including needle decompression, leg fasciotomy, and cricothyrotomy. Their results showed that the system enabled both novices and experts to communicate with increased precision, accuracy, and clarity. Novices were even able to complete the procedures they had never performed before. Fidalgo et al.~\cite{10049704} documented the expansive use of MR across other domains such as navigation, entertainment, education, and more through a large-scale survey. Their survey highlighted MR's critical role in improving performance and learning outcomes through remote assistance and training. 
The growing number of applications for collaborative MR calls for a need to understand the security risks associated with the mass adoption of MR devices in the industry and the readiness of users concerning them.

\textbf{Threats in MR:} Mixed Reality (MR) presents distinct security and privacy challenges arising from its integration of virtual and physical environments. Slocum et al.~\cite{slocum2024doesnt} highlighted vulnerabilities specific to Augmented Reality (AR) systems such as susceptibility to read and write attacks. Such attacks exploit AR's reliance on visual input and GPS data. These system components, often used to verify physical presence, can be spoofed leading to significant security breaches. Similarly, a study by Zhang et al.~\cite{zhang2023s} exposed side channels in MR sensor data to recover hand gestures, voice commands, and keystrokes from MR devices. They were able to identify launched applications with over 90\% accuracy, demonstrating that MR systems are vulnerable to security attacks.

Additionally, Cheng et al.~\cite{chenguser} provided an in-depth analysis of AR User Interface (UI) properties with significant security implications, showcasing empirical analyses across various platforms including ARCore, ARKit, Hololens, Oculus, and WebXR. These properties include the transparency/invisibility of virtual objects, the occupation of the same space by different objects, and the ability to add synthetic user input in MR-based applications. Understanding these AR UI properties and their security implications informed the design of our implemented attacks.

{\textbf{Threats in Collaborative MR:}}\label{RW:Threats} As the application of collaborative MR grows, so does the complexity of its security challenges ~\cite{paneva2024privacy, gallardo2023speculative}. Ruth et al.~\cite{ruth2019secure} focus on secure and private content sharing in multi-user environments, which is crucial for effective collaboration. Their work introduces a system called ShareAR with a control module designed to manage and secure how AR content is shared and interacted with. ShareAR helps mitigate one aspect of security threats associated with collaborative MR applications. Happa et al.~\cite{10.3389/fict.2019.00005} further elaborated on the vulnerabilities within network architectures that support collaborative MR, discussing the potential social, monetary, and psychological impacts of such exploits.

Additionally, Rajaram et al.~\cite{rajaram2023eliciting} highlight the growing use of AR in collaborative settings and the associated risks of data breaches, unauthorized access, and privacy violations. They conducted a user study to explore concerns and preferences about security and privacy in shared AR environments, using the findings to propose techniques such as granular access controls, encryption, and feedback mechanisms to address vulnerabilities.

Prior work on the variety of threats associated with collaborative MR settings encourages our focus on MR-specific attacks. Additionally, it presses on the need to investigate if users' unfamiliarity with the MR environment decreases the effectiveness of secure habits adopted by users. Our study aims to provide insights into how users interact with and perceive security threats unique to remote collaborative MR environments.

\begin{table}[!t]
\caption{Comparison of related works on MR security attacks.} 
\vspace{-2mm}\label{tab:rw}
\begin{tabular}{|l|P{1.75cm}|P{1cm}|P{1.5cm}|}
\hline
\textbf{Related Works} & \textbf{Collaborative} & \textbf{Remote} & \textbf{Deployed Attacks} \\
\hline
Cheng et al.~\cite{10.5555/3620237.3620289} & \xmark & \xmark & \cmark \\
\hline
Cheng et al.~\cite{chenguser} & \xmark & \xmark & \cmark \\
\hline
Zhang et al.~\cite{zhang2023s} & \xmark & \xmark & \cmark \\
\hline
Slocum et al. \cite{slocum2024doesnt} & \xmark & \cmark & \cmark \\
\hline
Lebeck et al.~\cite{lebeck2018towards} & \cmark & \xmark & \xmark \\
\hline
Happa et al.~\cite{10.3389/fict.2019.00005} & \cmark & \cmark & \xmark \\
\hline
\textbf{Our Work} & \cmark & \cmark & \cmark \\
\hline
\end{tabular}
\end{table}

{\textbf{User Perspective on Security Issues:}}\label{RW:User-Perspective}
Understanding user reactions to security threats within MR environments is crucial for the technology's adoption  ~\cite{masood2019augmented,gugenheimer2022novel, abraham2024you, o2023privacy}. Cheng et al.~\cite{10.5555/3620237.3620289} studied user responses to perceptual manipulation attacks. Their findings showed physical and behavioral reactions when users were unknowingly subjected to these threats. However, their study was not set in a collaborative context and participants were not aware that they were being subjected to security attacks. Additionally, their attack implementation relied on stimuli external to the MR environment. In contrast, our study simulates attacks in a remote collaborative MR environment where participants are explicitly informed that attacks may be present and are tasked with identifying and mitigating them. This approach aims to capture the level of difficulty participants face in recognizing and addressing attacks within MR environments. Similar research by Erickson et al.~\cite{Erickson2020SharingGR} explored the identification and impact of errors in remotely shared MR data. They emphasized the users' perspective by documenting how gaze data inaccuracies affected task performance, user trust, and subjective experience in MR settings. 

Lebeck et al.~\cite{lebeck2018towards} recognized the need to explore the security and privacy of MR users in the context of multi-user applications. Their study investigated users’ concerns when engaging with immersive AR technologies and found that users may treat virtual objects as real. This confusion between the real and virtual world can lead to risks such as deceptive virtual objects and identifies the need for access control in shared spaces. Their study primarily focused on co-located applications without implementing actual attacks, it serves as a motivation for our research to go beyond this scope. While these related works stress the growing interest and importance of secure user interactions within MR, they do not specifically analyze how users respond to different types of security attacks in a remote collaborative MR setting, leaving a gap in the literature,(see Table~\ref{tab:rw}). We address this gap by implementing such attacks and evaluating user perceptions and responses.

\section{Methodology}

In this section, we state the research questions motivating our studies and our threat model. We describe the four implemented attacks\,(\S3.2), our study design\,(\S3.3), pilot study findings\,(\S3.4), and main study protocol\,(\S3.5). Our user study design addresses the following research questions:

\begin{itemize}
    \item {\textbf{RQ1:} How do users’ concerns towards security attacks change when exposed to attacks that are exclusive to remote collaborative MR environments? }

    \item {\textbf{RQ2:} How do users perform in recognizing security attacks in remote collaborative MR environments and what are their mitigation strategies?}

    \item {\textbf{RQ3:} How are user experience and collaboration affected in the presence of MR security attacks?}
\end{itemize}

\subsection{Threat Model}
Our threat model considers multiple users interacting within a networked remote collaborative MR system. We assume that the MR application allows third-party integration, similar to prior studies~\cite{zhang2023s,chenguser, 10.5555/3620237.3620289}. This assumption draws on existing platforms, such as VRChat~\cite{vrchat}, which enable third-party integrations through Open Sound Control~\cite{opensoundcontrol}, granting access to sensor data. We anticipate that future large-scale MR platforms will adopt extensive third-party integrations, much like Alexa Skills in Internet-of-Things devices. The third-party integration code could compromise the security or functionality of the MR application. This risk is particularly concerning in remote collaborative MR environments, where 3D interactions and the shared virtual space are essential to the user experience. Malicious entities could exploit vulnerabilities to disrupt critical collaborations, manipulating users' perception of the environment, and impairing their ability to coordinate, potentially resulting in physical or psychological harm to users and bystanders.

The attacks demonstrated in this paper require only minimal information obtainable through third-party APIs. For example, the spatial occlusion attack can be executed using the head position of the user to infer the approximate area of object interaction. Similarly, click redirection and object occlusion attacks can be implemented by accessing a list of game objects in the scene without precise knowledge of their spatial positions or task relevance. Lastly, the latency attack operates independently of the headset environment and instead exploits network vulnerabilities.

\subsection{Attacks Implemented} \label{attacks}
In order to investigate user behavior around various security attacks, we created a prototype based on ``SurfShare'' developed in~\cite{10.1145/3631418}. This prototype is implemented in Unity3D version 2019.4.38f1 on the Microsoft HoloLens 2. The prototype allows two Microsoft HoloLens 2 devices to exchange a part of their physical environment, through a portal, over the network. The portals in the application allow users to create virtual 3D representations of physical objects in their environment and stream them over to be viewed and manipulated by all users in the session. Each 3D representation has a ``handle'' attached to it, which can be used to resize the object. This handle can be seen as a white box on top of each object as shown in Figures~\ref{fig:teaser} and~\ref{fig:combined_figures}. 

We implemented four types of attacks, including two inspired by~\cite{chenguser}. Based on whether similar types of attacks exist in 2D computing environments, Latency and Click Redirection attacks are classified as ``familiar'' attacks, while Object and Spatial Occlusion are considered ``unfamiliar''.

\textbf{Latency Attack:} Latency attacks are defined as a malicious increased delay in a communication channel aimed to stop applications from responding to the requests within a reasonable time~\cite{chen2024overload}. We apply the same concept in the remote collaborative MR environment where network communication synchronizes the orientation and interactions of shared virtual objects between users. We used the Mirror package in Unity3D to connect the MR headsets through a desktop server machine and manipulate network traffic. We used Mirror's built-in latency simulator to implement the latency attack. We set the packet loss rate to 0.5\% and added a variable latency of around 10 ms in the network. These parameters were determined by experimenting with various values to find the point where user experience is significantly affected without causing the application to become unresponsive.

\textbf{Click Redirection Attack:} This attack is inspired by the web-based attack called clickjacking. Traditionally, this attack refers to tricking the victim into clicking an element that the victim never intended to click. This is usually done by making the element barely visible or completely hidden~\cite{saini2019you}. Previous work by Cheng et al.~\cite{chenguser} implemented a clickjacking attack in an MR environment by placing two virtual objects in the same position in 3D space. They exploited the inconsistency in rendering and interactivity order of those two objects to implement this attack. 

In our study, we introduce a novel attack called ``Click Redirection." Unlike the traditional web-based attack, both objects are visible in this attack, and unlike MR clickjacking, both objects can occupy different positions in the environment. We designed the attack such that when a user tries to move one virtual object via interaction, another object will move instead, and vice versa. This is illustrated in Figures~\ref{fig:sub1} and~\ref{fig:sub2}. Implementing this attack in a 3D immersive environment also means that when the victim tries to move one attacked object, they will not witness the movement of the proxy object in real-time if it is outside their field of view. However, they will still notice the changed position of the proxy object when they look around in their environment. Furthermore, the collaborative environment can make the unintended movement of virtual objects a potential cause of mistrust and confusion between the collaborators.

To implement the click redirection attack, we loop through the virtual objects in the 3D environment and randomly select two objects: an attacked object and a proxy object. The attack deletes the attacked object's 3D transform manipulation code and replaces it with that of the proxy object. The attack effectively disables manipulation of the attacked object and applies all manipulations to the proxy object instead.  
The attack waits until a minimum of four objects have been created and shared with collaborators before click redirection is activated. 
    
\textbf{Object Occlusion Attack:} Cheng et al.~\cite{chenguser} identify invisibility/transparency as one of the UI properties of MR systems that can be exploited to implement security attacks on MR devices. They used this property to implement a Denial-of-Service (DoS) attack by making a transparent/invisible virtual object overlay on a visible virtual object. This transparent/invisible object blocked all interactions with the underlying visible virtual object. While our implementation of the object occlusion attack shares similarities with the DoS attack discussed in~\cite{chenguser}, we chose to call it the ``Object Occlusion Attack'' for two main reasons. First, we wanted to avoid confusion with the more conventional definition of DoS attacks that involve overwhelming a system with network load ~\cite{mitrokotsa2007denial}. Second, our attack was different from the DoS attack in~\cite{chenguser} since under our attack, one could still move an object in our environment if they got close enough to be within the invisible bounding box around the attacked object.

Our attack waits for virtual objects to be created in the environment. Once the minimum threshold of four virtual objects is reached, a transparent/invisible bounding box encapsulates a random shared virtual object in the environment. When a user tries to interact with the visible object from a distance using the ray cast, the invisible bounding box blocks the ray cast and does not let it reach the visible object as shown in Figure~\ref{fig:sub4}. The transparent bounding box follows the object such that the object always remains within its bounds. This makes it impossible to interact with the attacked object via ray cast interaction and forces the near-interaction gesture in the case of Microsoft Hololens 2. 

\textbf{Spatial Occlusion Attack:} This attack is similar to the object occlusion attack described above. However, as the name suggests, the purpose of this attack is not to occlude a particular visible object, but a specific region in the environment. The occlusion ensures that the user cannot interact with any object in the occluded space unless they physically reach inside the bounds of the transparent/invisible bounding box. The transparent/invisible bounding box has been implemented differently for this attack in the following ways: First, the bounding box is larger, allowing multiple virtual objects to be encapsulated within its bounds, as shown in Figure~\ref{fig:sub3}. This contrasts with the object occlusion attack, where the transparent/invisible bounding box resizes to encapsulate only the target object. Second, it remains static in space and objects can be released from the attack if they are moved out of the bounding box via near-interaction. Similar to other attacks, we make sure the minimum threshold of four objects is met before launching the attack. We also initialize the position of the box such that it encapsulates at least one random shared object in the environment.

\subsection{Study Design} \label{study_design}

\begin{figure}[!t]
    \centering
    \begin{subfigure}[b]{0.3\linewidth}  
        \centering
        \includegraphics[width=\linewidth, height=3cm]{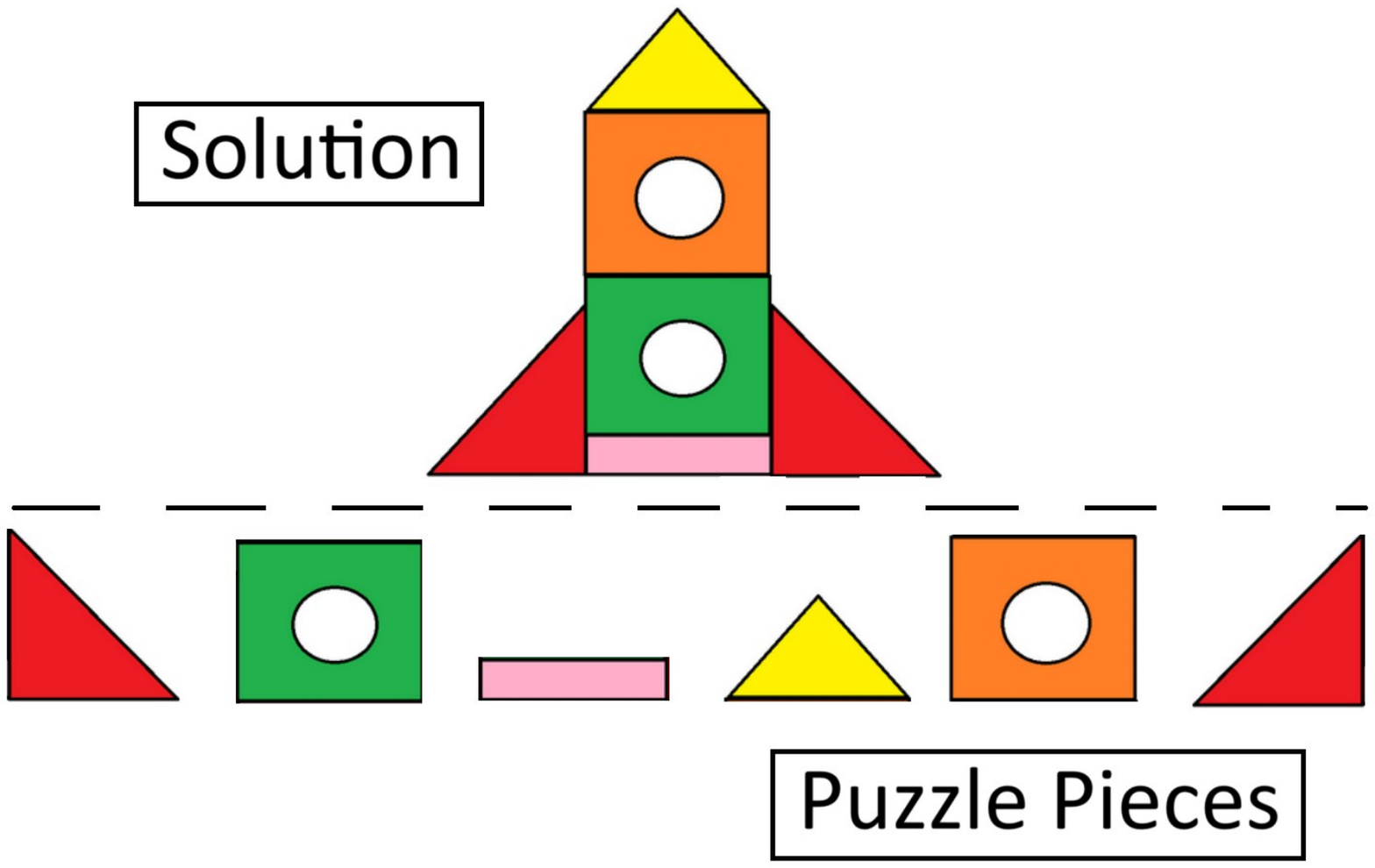}
        \caption{}
        \label{combined_figures:sub1}
    \end{subfigure}
    \hspace{0.03\linewidth}  
    \begin{subfigure}[b]{0.3\linewidth}
        \centering
        \includegraphics[width=\linewidth, height=3cm]{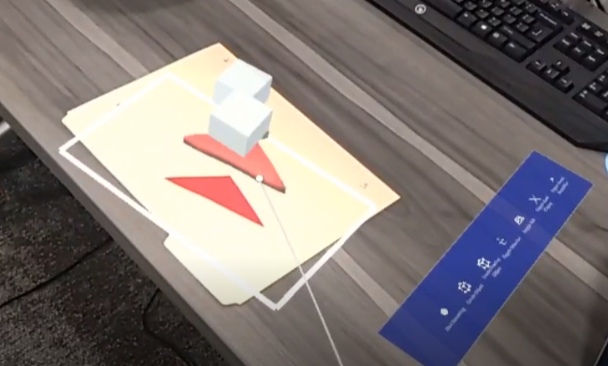}
        \caption{}
        \label{combined_figures:sub3}
    \end{subfigure}
    \hspace{0.03\linewidth}  
    \begin{subfigure}[b]{0.3\linewidth}
        \centering
        \includegraphics[width=\linewidth, height=3cm]{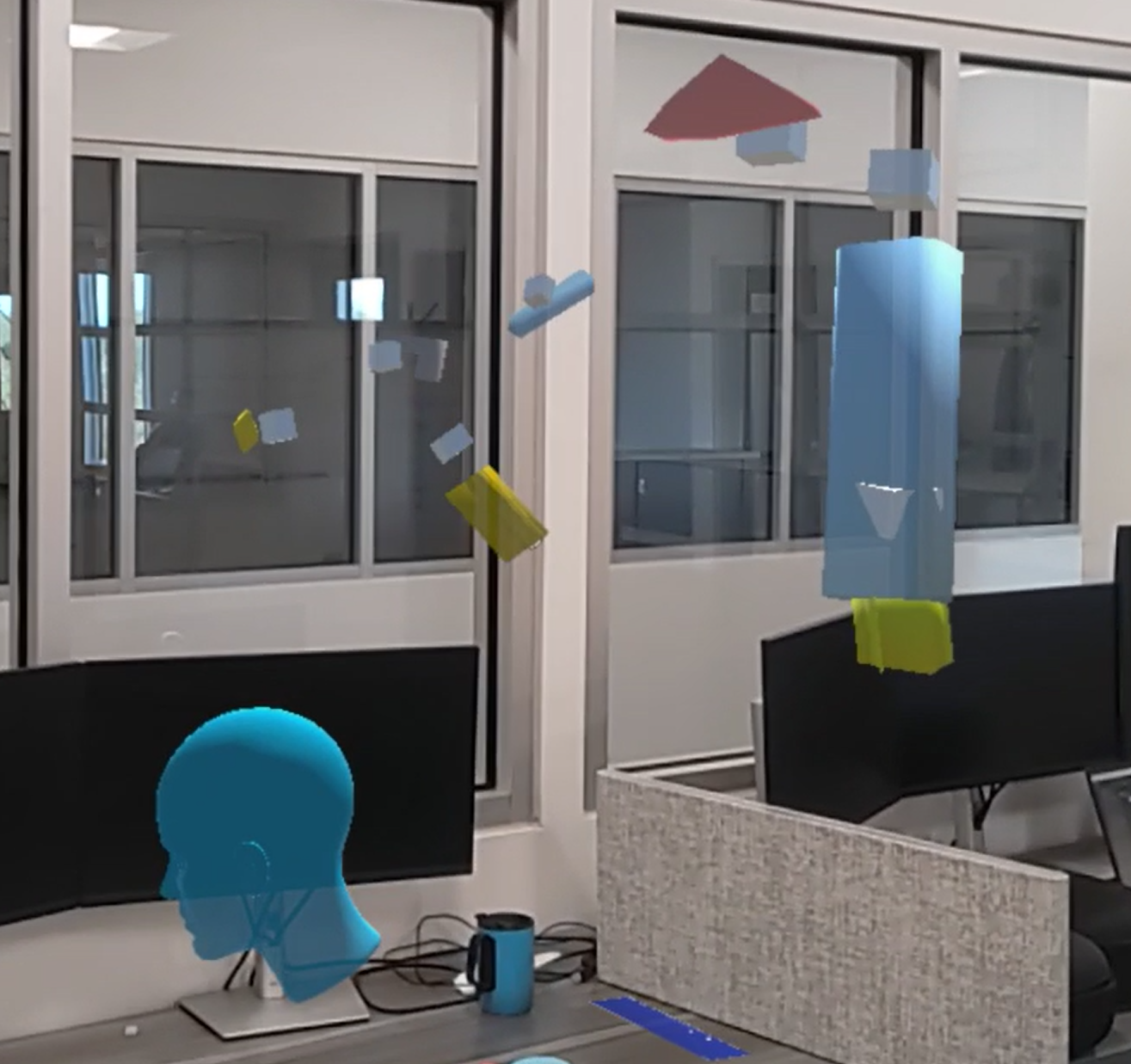}
        \caption{}
        \label{combined_figures:sub4}
    \end{subfigure}

    \caption{(a) The target puzzle solution and provided shapes. (b) Participants place the 2D cutouts within their portal to create and share as a 3D virtual object. (c) Partially completed tower in the MR environment.}
    \label{fig:combined_figures}
\end{figure}

\textbf{Task Design:} The task was designed to be collaborative, remote, and representative of real-life applications with spatial and temporal coordination. The task was inspired by prototyping and training guidance for assembly. We provided each participant with paper cutouts representing various shapes. Participants were asked to convert these cutouts into virtual objects within the MR environment. They then either stacked the shapes on top of each other within a set time limit or used them to recreate a sample image provided as part of the task (Figure~\ref{combined_figures:sub1}).

\textbf{Study Protocol:} Our study started with a pair of participants present in the same room. The participants received the consent form to read and sign. After this, they were introduced to the MR device (Microsoft HoloLens 2) using the built-in Tips application, which provides a tutorial on how to operate the HoloLens~2.

Next, the participants were briefed about the user study procedure through a PowerPoint presentation. This presentation included an introduction and demo of our MR application components and a description of the assembly task. The presentation also primed participants to think about security attacks through two slides listing the names and definitions of traditional cybersecurity attacks. Priming participants was essential to establish a consistent baseline understanding of common security attacks before experiencing any attacks in MR. Prior studies \cite{chenguser} show that participants often suspected the premise of the study due to the known research focus on security from the experimenters' lab, which unintentionally introduced variability in responses. Explicit priming of participants not only addressed this issue but also aligned the study with realistic scenarios, such as security training provided to new employees in organizational contexts.

After the presentation, an acclimatization round of the task was conducted without attacks; during this round, we guided the participants through task completion. Once the participants felt comfortable with the device, they were taken to separate rooms for the data collection. Participants then filled out the pre-task questionnaire. At the start of each condition, Participant~1 used guiding marks on the table to position a square representing their portal. Participant~2 then placed the representation of that portal on their table, as a reference to Participant~1's position in their environment. Next, they moved on to position their portal using another guiding mark on a table. After configuring the portals, one of the participants enabled the remote connection by clicking the ``Start Streaming" button in the menu panel. This action established an audio connection and enabled participants to see each other's virtual avatar representations within their respective environments. Once this connection was established, researchers started the timer and both the participants placed paper cutouts provided to them in the portal and created virtual 3D objects out of them using the ``Create Floating Object" menu button as shown in Figure~\ref{combined_figures:sub3}. After creating the virtual objects, participants moved them around in the environment (Figure~\ref{combined_figures:sub4}) to solve the target puzzle of the rocket shown in Figure~\ref{combined_figures:sub1}. 

Participants repeated the task under baseline and attack conditions. Both participants experienced the same type of attack during each round. We recorded participants' interactions from a first-person perspective during the task. However, to avoid degradation of performance in HoloLens 2 and to maintain user privacy, we captured the generic holographic view without the raw camera feed. We also noted the time it took for participants to complete the task in each condition. After each round, a questionnaire was administered to understand changes in participants' perceptions. The study concluded with a semi-structured interview to understand participants' experiences at a deeper level. 

\subsection{Pilot Studies} \label{Pilot Studies}
We ran two sets of pilot studies before finalizing the study design.

\textbf{Pilot Study 1:}
The first Pilot Study consisted of four user studies with eight total participants. We implemented three types of attacks in the prototype including latency attack, object occlusion, and spatial occlusion. While the baseline condition was always the first condition, the order of attack conditions was randomized for each user study.
 
 The results of Pilot Study~1 revealed a pattern of differences in user behaviors towards various attacks. For example, attacks unique to MR were more frequently misattributed. The pilot study also provided us with valuable feedback on our study design. First, the latency attack was not the best choice for a familiar attack since the degradation in user experience brought the risk of high frustration in participants. We believe this could affect their willingness to provide detailed answers to the subjective questions. Moreover, the participants had no direct means to mitigate the latency attack. Second, the results revealed that most participants were unable to notice any differences between the two occlusion attacks, and both conditions produced very similar results. We also observed that since the baseline condition was always the first condition, the average usability scores of participants in the baseline round ($3.05\pm0.75$) were worse than the attack conditions ($3.2\pm0.81$, $3.22\pm0.92$, and $66\pm7.5$). This was unexpected and motivated us to randomize the order for the baseline condition in future studies. 
 
 Another insight from Pilot Study 1 was related to participants' focus on attack conditions. We observed that participants were too concerned about completing the task on time to notice the attacks in the environment. This made us realize that we need to modify the task by removing time pressure to allow participants to think about the attacks. Furthermore, each participant had to create a portal in their virtual environment that required three pinch gestures to mark the three corners of the rectangle in a specific order. Most of the participants struggled with this method of portal creation, increasing fatigue and extending the duration of the study.

\textbf{Pilot Study 2:} The second Pilot Study also consisted of four user studies and eight different participants. We made two changes to the attack conditions: (1) We dropped the object occlusion attack and (2) replaced the latency attack with the click redirection attack. Therefore, the conditions for Pilot Study 2 were baseline, spatial occlusion attack, and click redirection attack. As mentioned above, the two occlusion attacks provided similar results so we excluded object occlusion to reduce the user study time and fatigue of participants. The reason for including the spatial occlusion attack was that similar examples are rare in 2D computing environments. Similarly, the results of Pilot Study 1 made a case for replacing the latency attack because we did not observe any mitigation strategies. As a result, we conceptualized and implemented the click redirection attack instead.

We also changed the study protocol so that participants were primed to think about security attacks through the task introduction presentation, and a semi-structured interview was added at the end to gain more meaningful insights about the participant's experience. Furthermore, we used a Latin square generator to decide the order of each condition between experiments, including the baseline condition. This helped reduce the ordering effect in our user studies. As a result, we witnessed results more consistent with our expectations, with the baseline condition having a higher average system usability rating ($3.88\pm 0.74$) compared to click redirection ($3.35\pm 0.75$) and spatial occlusion ($3.12 \pm 0.66$). 

The task was also modified from Pilot Study 1 to increase the level of engagement and standardization between each collaborating pair. Instead of asking participants to simply stack pieces to build a tower, we decided to provide them with a sample image to replicate as shown in Figure~\ref{combined_figures:sub1}. Additionally, we decided to remove the time restriction and simply measure the time it takes for participants to build the rocket as a metric of performance. Pilot Study 2 yielded promising results, providing a strong foundation for proceeding with the main study.

\subsection{Main Study} \label{user-study}

We recruited 20 participants divided into 10 user study sessions for the user study through university listservs. Thirteen participants were male and seven were female. The age of participants ranged between 19 and 37. All participants were Virginia Tech students. We asked participants about their prior experience with MR devices and found that six participants had no prior experience, while fourteen participants had limited experience, having used an MR device a few times. We countered for that lack of experience through the acclimatization round before the start of data collection. All participants were at least 18 years old, had normal vision without glasses, and were English speakers. Virginia Tech's Institutional Review Board approved this and all pilot studies. Participants signed consent forms before the start of each study and were compensated for their participation with a cash amount of \$20.

We included two types of attacks in the application for the final study: click redirection and spatial occlusion. The attacked objects were selected randomly for each participant. This means that not only were the objects attacked for each of these conditions different for both participants in each round but they also differed between rounds. 

In this study, we aimed to address our research questions (\textbf{RQs}) by testing corresponding hypotheses. We hypothesized that user's concerns about security attacks would increase significantly under threats unique to remote collaborative MR environments (\textbf{RQ1}); users would perform significantly worse at identifying such attacks (\textbf{RQ2}); and user experience would significantly degrade while collaboration increases as users rely on one another under attack (\textbf{RQ3}). The reasoning behind our hypotheses stems from the lack of familiarity with MR-specific attacks.

\section{Results}
In this section, we analyze and draw conclusions from the quantitative data (\S\ref{R:Quantitative}), observational data (\S\ref{R:Observational}), and subjective data (\S\ref{R:Qualitative}) collected during our study.

\subsection{User Reported Quantitative Measures} \label{R:Quantitative}

Participants filled out a post-task questionnaire after every condition including standard scales regarding their perceived task workload, system usability, and security perceptions. The questionnaire consisted of five parts: (a) NASA TLX questionnaire~\cite{tlx}, (b) System Usability Scale (SUS)~\cite{article}, (c) subset of questions from the Mixed Reality Concerns (MRC) Questionnaire~\cite{mrc}, (d) participant's rating of their own performance ``How would you rate your performance in the last session?", and (e) an open-ended question asking the participants to report their experience with the attack ``Based on your prior knowledge, do you think you were under a security compromise in this condition? Try to name the kind of security compromise if possible." Questions from (a) - (d) were asked on a 5-point Likert scale. Table~\ref{tab:questions} lists the included MRC questions. Please see the Supplementary Material for a list of all prompts in the questionnaire.

\begin{table}[ht]

\centering
\begin{tabular}{|>{\centering\arraybackslash}p{8cm}|}
\hline

\multicolumn{1}{|c|}{\textbf{Subset of Mixed Reality Concerns (MRC) Questionnaire}} \\ \hline
I am concerned about the possibility of non-authenticated individuals gaining access to this MR system. \\ \hline
I am concerned about the potential of this MR system to influence my behaviors in ways that could be detrimental to my well-being.\\ \hline
I am sure that this MR system is maintaining a secure environment. \\ \hline

\end{tabular}
\caption{MRC questions included in the post-condition questionnaire.}
\label{tab:questions}
\end{table}

We compared participants' answers to these questions across the three conditions (baseline, spatial occlusion attack, and click redirection attack) and reported the results below. These quantitative results aim to answer \textbf{RQ1}. We calculated the unweighted score of each questionnaire by summing participants' ratings on a 5-point Likert scale. The rating for the subscale measuring success in NASA TLX was inverted in the final calculation to align with the other subscales, ensuring that higher scores consistently represent a higher task load. We took the average score of all participants to test for significant main effects. We used the Shapiro-Wilk test to test our data for normality. Our data did not meet the normality assumptions so we used Friedman's test to identify the significance of differences between conditions for each scale. If we found a $p < 0.05$, we ran the Wilcoxon Signed-Rank test as the posthoc test with Bonferonni correction between each pair of conditions. The Supplementary Document presents all of our statistical results.

The average score for task load conditions did not show any practical difference. The average task load for baseline was $14.20\pm 3.39$, click redirection was $14.70\pm3.03$ and spatial occlusion was $13.95\pm2.98$. These scores were calculated out of 30 with each of the 6 subscales contributing a rating out of 5. We also compared the ratings for subscales of the NASA TLX questionnaire and noticed that the average rating for frustration level was considerably higher in the click redirection condition $(2.30\pm 1.08)$ compared to the spatial occlusion condition $(1.06\pm 1.04)$. This difference was near significance $(p = 0.07)$, with a large effect size $(r = - 0.718)$. In other words, the frustration level shifted from not at all to slightly frustrated on average which indicates increased noticeability of frustration amongst participants. The comparison of average scores for SUS did not show any practical effect either. The baseline condition had an average score of $17.2\pm 3.72$, the spatial occlusion condition had an average score of $17.7\pm3.88$, and click redirection had an average score of $16.3\pm4.14$. These scores were calculated out of 25 with each of the 5 subscales contributing a rating out of 5.

\begin{figure*}[ht]
  \centering
  \begin{subfigure}{0.32\linewidth}
    \centering
    \includegraphics[width=\linewidth]{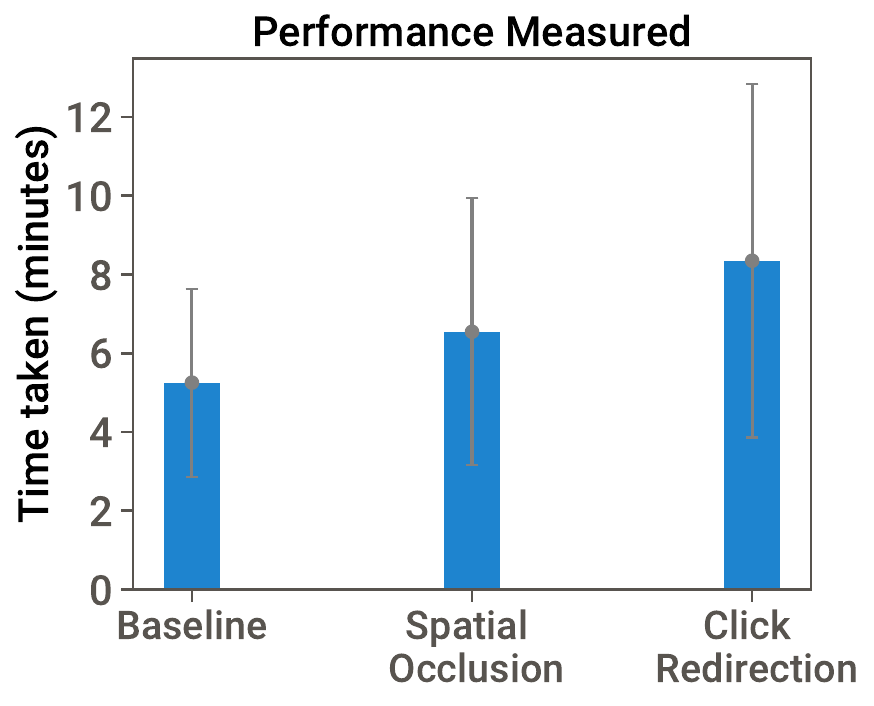}
    \caption{}
    \label{fig:performance_metrics}
  \end{subfigure}
  \hfill
  \begin{subfigure}{0.32\linewidth}
    \centering
    \includegraphics[width=\linewidth]{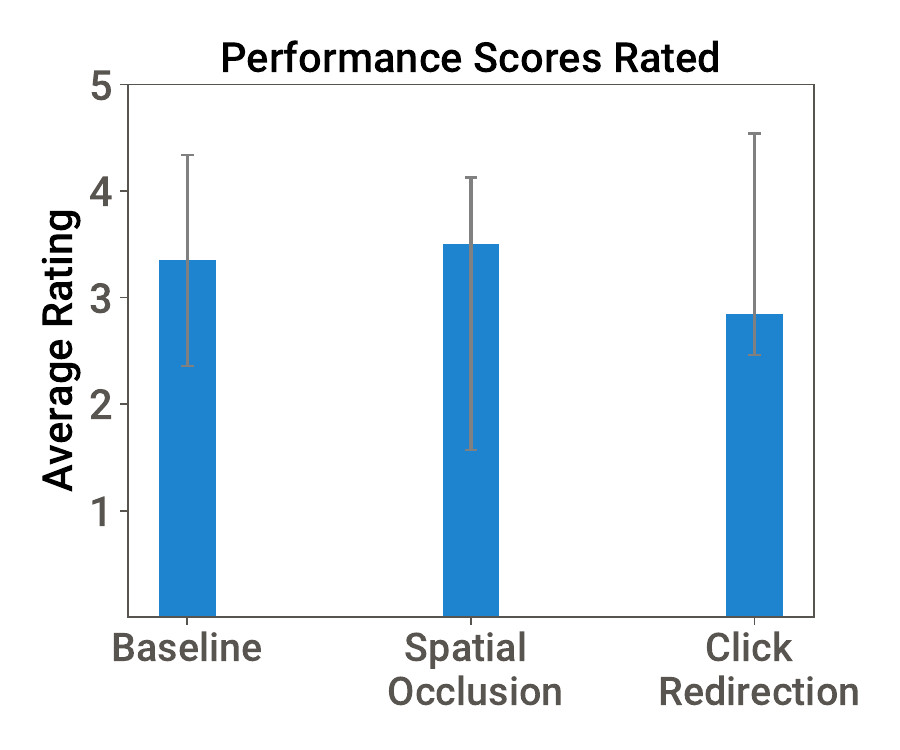}
    \caption{}
    \label{fig:performance_rated}
  \end{subfigure}
  \hfill
  \begin{subfigure}{0.32\linewidth}
    \centering
    \includegraphics[width=\linewidth]{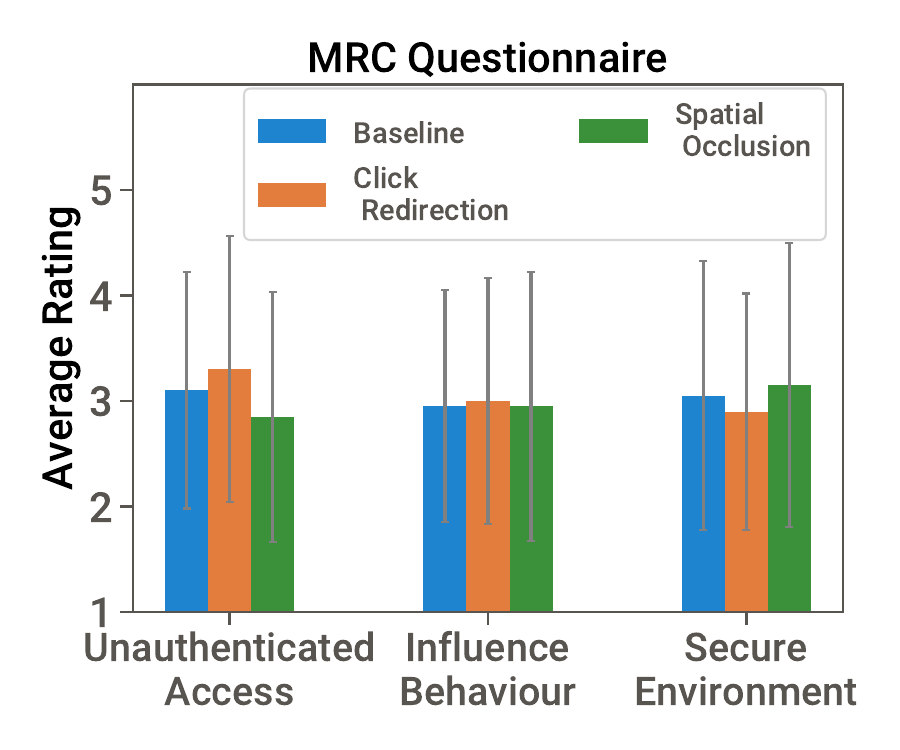}
     \caption{}
    \label{fig:MRC}
  \end{subfigure}

  \caption{(a) Average time taken by participants across conditions. (b) Participants' average ratings of performance, showing higher ratings for spatial occlusion despite worse performance. (c) Responses to security concerns from the MRC Questionnaire mapped by attack condition.}
  \label{fig:combined_full_width}
\end{figure*}

Participants' responses to the MRC Questionnaire indicated that different types of attack conditions raise different types of security concerns as shown in Figure~\ref{fig:MRC}. The average rating for concern regarding unauthenticated access was higher for the click redirection condition ($3.30\pm1.26$) than that for the spatial occlusion condition ($2.85\pm1.18$). However, the average rating for concern regarding the security of the environment was higher for the spatial occlusion condition ($3.15\pm1.34$) than the click redirection condition($2.90\pm1.12$). The level of concern generally remained low across conditions. 

The average rating of participants' perception of their performance was also compared across conditions. The results indicated that participants exhibited the highest confidence in their performance during the spatial occlusion condition $3.50\pm1.28$, followed by the baseline condition $3.35\pm0.98$, and the lowest confidence during the click redirection condition $2.85\pm1.03$. Although this difference is not significant, this is an interesting finding because when we measured their actual performance in terms of time taken to complete the task, we found that the average performance during the spatial occlusion condition was worse than the average performance during the baseline condition as shown in Figure~\ref{fig:performance_metrics}. On average, the task took $5.25\pm2.39$ minutes to complete in the baseline condition, compared to $6.54\pm3.39$ minutes in the spatial occlusion condition. However, the click redirection attack did lead to the longest task completion time, averaging $8.54\pm4.48$ minutes. The difference in performance time for click redirection condition and baseline was found to be statistically significant $(p = 0.02)$ with a large effect size $(r = -0.956)$.

\subsection{Observed Measures} \label{R:Observational}
As mentioned in \S\ref{study_design}, we recorded the user's first-person perspective while performing the task. We watched each video and prepared summary notes for each session. The themes focused on the way users faced the attack, how users reacted to the attack, how users mitigated the attack, and what kind of collaborative effort was witnessed during the attack. We also cross-referenced our observations with the open-ended question in the post-task questionnaire where participants were asked to self-report these measures. After collecting summary notes for each session, we grouped them in terms of attacks recognized and mitigation techniques against these attacks to answer \textbf{RQ2}. 

\begin{figure}[h]
  \centering
  \includegraphics[width=\linewidth]{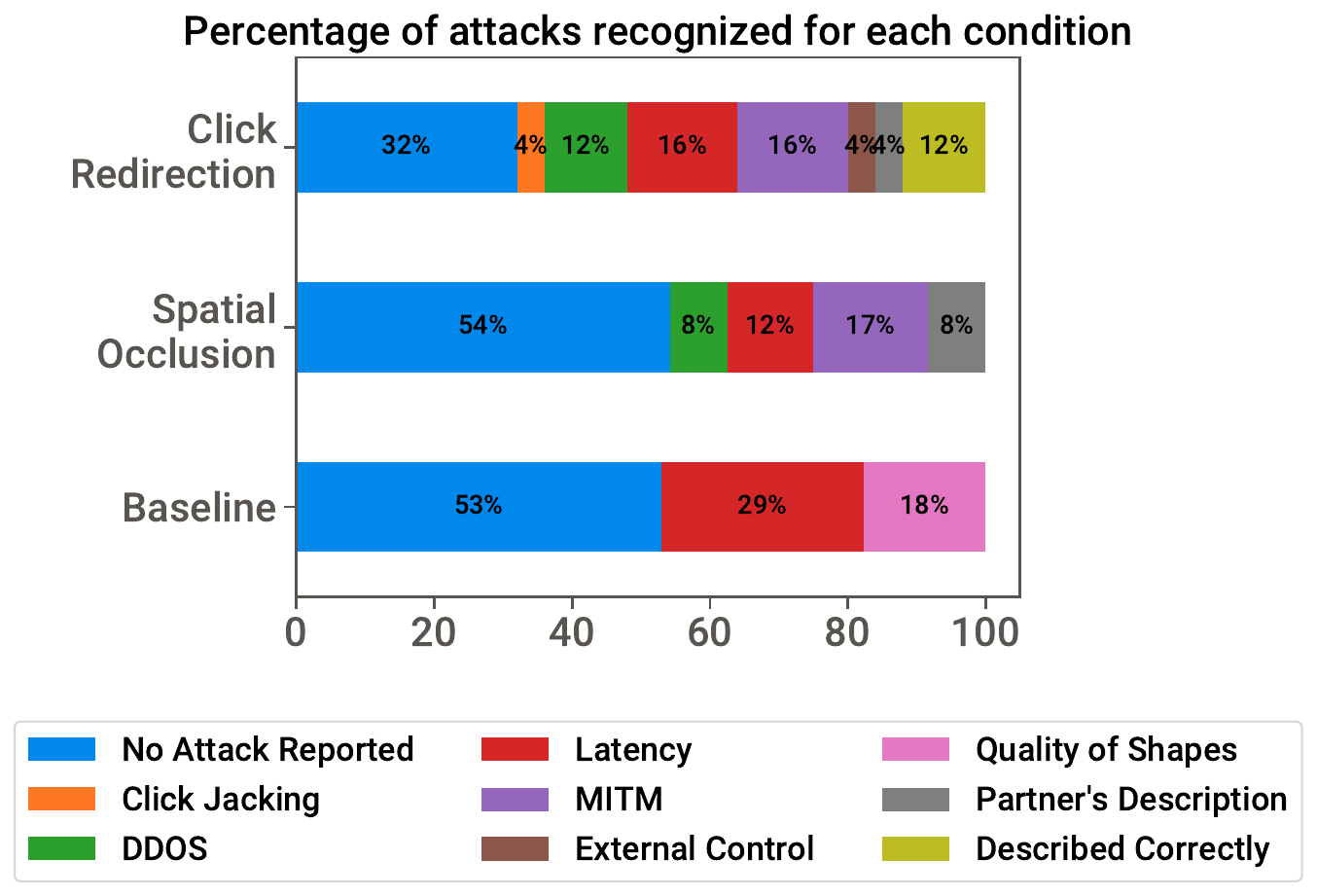}
  \caption{This graph shows the distribution of types of attacks recognized for each condition. There is low recognition and high misclassification by the participants for the implemented attacks.}
   \label{fig:recognition}
\end{figure}

Figure~\ref{fig:recognition} shows the distribution of types of attacks identified for each condition. We noticed that a majority of the participants reported no attack in the baseline condition, and for the 30\% who did report an attack, it was a result of inaccurate pinch detection that prevented the user from selecting the desired object. We witnessed a wide variety of recognition patterns and attack labels for click redirection and spatial occlusion conditions. Furthermore, we observed that over 50\% of participants reported no attack during the spatial occlusion condition, while only about 30\% suspected no attack during the click redirection condition. This shows a pattern of higher attack recognition in the familiar, click redirection condition compared to the unfamiliar, spatial occlusion condition. Another interesting finding was that 12\% of the attacks recognized for click redirection involved describing the attack accurately without labeling it in terms of the standard list of web-based cyber-attacks provided to them. For instance, User 1 of User Study 2 described the click redirection attack in the following words, \textit{``I made three objects, and while selecting one of them, when I tried to drag it, the other object was getting moved."} However, participants did not provide such descriptions for the spatial occlusion attack.

\begin{figure}[h]
    \centering
    \includegraphics[width=\linewidth]{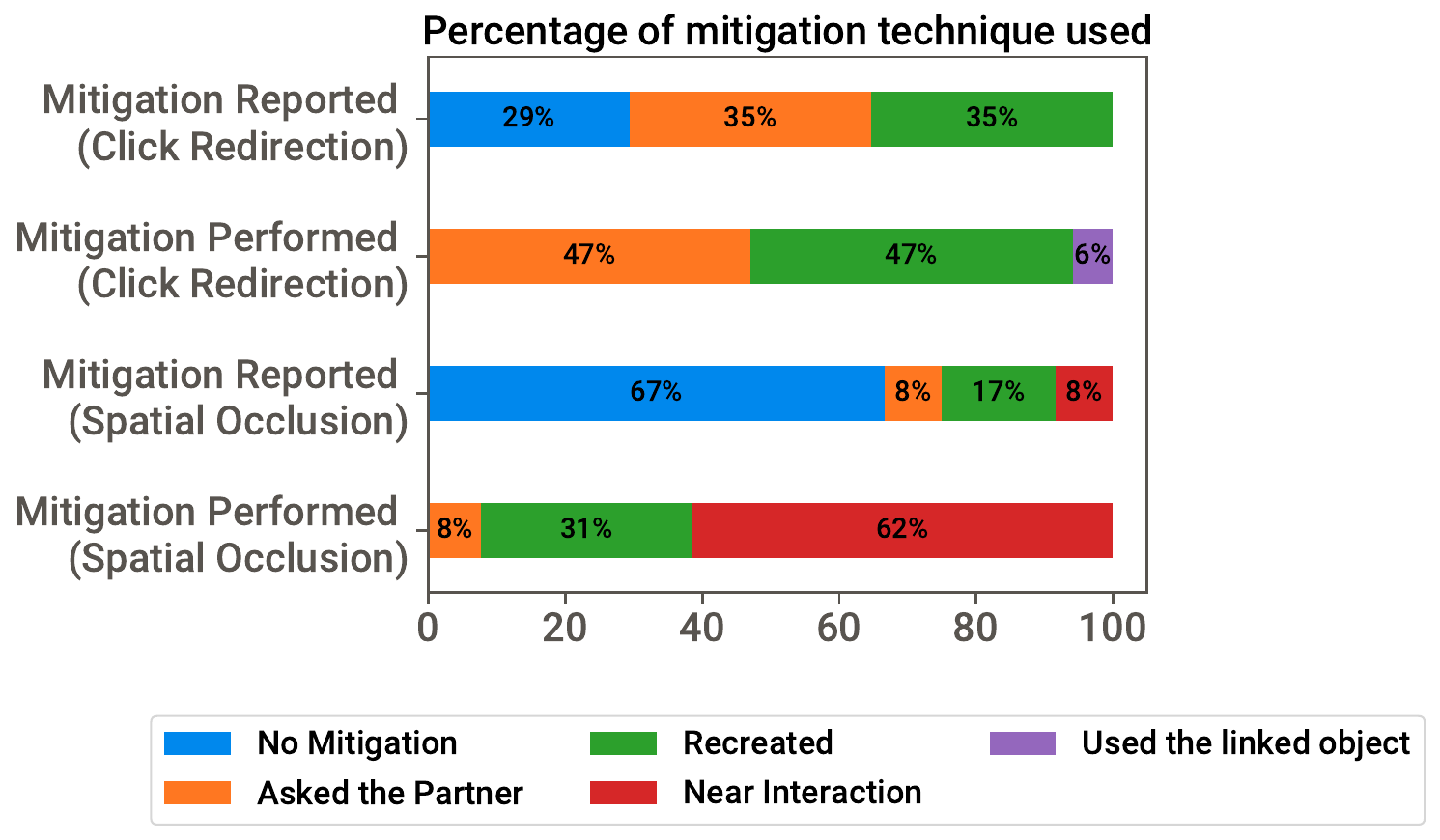}
    \label{mitigation_Spatial}

  \caption{ Distribution of mitigation techniques reported and performed for different attack conditions. There is a discrepancy in the mitigation techniques reported and performed by the participants for both the spatial occlusion and click redirection attacks.}
  
  \label{fig:mitigation}
  
\end{figure}

Similar to attack recognition patterns, user mitigation techniques were also grouped by theme. However, in this case, we did this grouping twice. Once for the mitigation techniques that we observed in the recordings and the other for the mitigation techniques that the participants reported. There are five possible mitigation techniques for both attacks (No mitigation, Ask your partner to move the attacked object for you, Recreate the virtual object, Use near-interaction to approach the object and Use the proxy object to move the intended object). Three types of behaviors are common between the two attacks: recreating the virtual object, asking their partner to manipulate the attacked object instead, and not mitigating the attack at all. Two mitigation techniques are attack-specific, one for each condition. In the case of click redirection, the attack-specific mitigation technique required participants to identify the attacked and proxy objects and use one to move the other in the intended place. Whereas, for spatial occlusion attack participants needed to physically approach the virtual object and only move it via near-interaction. Figure \ref{fig:mitigation} shows the distribution of each mitigation technique used and reported across both attack conditions. 

We noticed a discrepancy between the mitigation techniques reported by the participants and the actual mitigation techniques they used in the conditions. For the click redirection condition, most participants correctly reported the mitigation technique they used. However, around 30\% of the participants never reported a mitigation technique despite performing mitigation during the task. The results for the spatial occlusion attack had an even larger discrepancy between the data from the recording and participants' self-reporting. Around 67\% participants failed to report their mitigation technique. Most of these participants (61\%) had employed near-interaction to mitigate the attack. However, only one participant reported using this technique to complete the task in the presence of the spatial occlusion attack. 

\subsection{Subjective Analysis} \label{R:Qualitative}

\textbf{Attack Recognition and Attribution:} \label{Attack Recognition and Attribution}
We determined that the baseline condition was an effective control, as the majority of participants reported no perception of security attacks. However, users often misclassified the type of attack they experienced. User 1 from User Study 7 said about the click redirection attack, 
\textit{``I was not able to move some shapes while my partner was not able to move some other shapes. I suspect this to be due to Denial-of-Service (DoS) or Man-in-the-Middle (MITM) attack."} Participants frequently misidentified the spatial occlusion attack as a system failure to respond to their actions, leading them to perceive it as a DoS or latency attack. In their interview, User 2 from User Study 1 described the situation as follows: \textit{``It was a DoS attack. The system was not handling my request."} User 2 from User Study 10 also reported, \textit{``There were delays in this [spatial occlusion] condition, this may be due to latency attack."}

The results also revealed the tendency of participants to attribute the issues they face, due to the attacks, to factors like user errors, technological glitches with the device network connectivity, and the overall performance of their partner. During the spatial occlusion condition, User 1 from User Study 6 thought that the attack was a technological glitch and complained to the investigator about a faulty application. During the interview the same participant reported, \textit{``I did not understand that objects had to be moved by going physically near, so that was my user error. I was probably not doing air tap right."} During another study, the proxy object was already a part of the partially completed puzzle solution that User 1 from User Study 5 was assembling when they encountered the click redirection attack. However, the proxy object was displaced from that arrangement when the participant attempted to move the attacked object. They noticed this displacement later as they looked back at their solution and said to their partner shockingly, \textit{``Oh my! Everything has changed. Please stop moving objects."} Their partner replied that they had not moved any objects out of the arrangement, to which User 1 of User Study 5 responded, \textit{``Just stop doing everything for now!"} convinced that their partner had moved the object.

We observed a general lack of suspicion among participants regarding the possibility of attacks despite being specifically instructed to watch for unusual behaviors in their environment that could indicate a security threat. User 2 of User Study 2 even reported that they felt none of the conditions involved an attack. During the interview, they suggested a need for user training in response to the issues faced during the task completion.

We also noticed that spatial occlusion was less frequently reported as a potential security threat in post-experiment interviews. Even when User 2 of User Study 8 was actively under the attack of spatial occlusion they did not tell their partner about the problem they were facing. Instead, they said, \emph{``Wait, let me figure something out."} We also believe users' understanding of learning effects may have masked attack perceptions in the earlier conditions. User 2 of User Study 5 was subjected to the conditions in the following order: spatial occlusion, click redirection, and baseline. In the post-experiment interview, this participant described that they felt their performance improved with each condition. They attributed difficulties in earlier conditions to their lack of familiarity with the prototype rather than the presence of attacks.

\textbf{Participants' Collaboration:} Our data analysis also revealed interesting behaviors within the collaboration of users in the presence of a security attack, as we explored \textbf{RQ3}. The type of impact varied between user studies. Sometimes, the attack led to increased collaboration where participants would communicate with their partner about the issue they were facing and rely on their help to resolve it. At other times, they would blame their partner for miscommunication and misplacement of objects in cases where the attack was not recognized (User 2 from User Study 5). Another case of a participant's mistrust due to an attack was witnessed in the post-experiment interview of User 2 from User Study 3 where the participant explained, ``\textit{It is easy to identify attacks when you're alone because you know if you see a change you did not make, it's a problem. But when you have a partner there are two suspects to select from, the attack and the partner."} In the case of User 1 from User Study 6 we noticed increased reliance on the partner when subjected to an attack. Once they realized they could not move an object as expected, they assumed the issue was on their end. From that point onward they just created the virtual objects out of the paper cutouts provided to them and asked their partner to move and arrange those objects to complete the puzzle. 

Participants' attack experience was different compared to their collaborators because we randomized which object or region was attacked for each individual. This asymmetric experience led to miscommunication and hindered collaboration. For example, User 2 from User Study 5 described their spatial occlusion attack to their partner as, \textit{“I can't reach the object, my ray is being blocked”}, hoping for help or guidance. However, since User 1 was not experiencing the same issue, they did not respond.

Even when both participants faced the attack at the same time we noticed varying levels of persistence towards mitigating the attack. User 2 from User Study 10 spent four minutes trying to move a virtual object targeted by the click redirection attack before recreating the object. User 1 from User Study 10 was also stuck trying to move the object simultaneously. User 2 advised User 1 to recreate the object from their own experience. However, User 1 continued attempting to move the attacked object, until User 2 finally moved the attacked object for them. Overall, we noticed that collaboration between partners increased more often in the case of click redirection attacks. On the contrary, most participants did not even mention their attack experience to their partners in case of the spatial occlusion attack.

\textbf{Perceptions about Real Life Applications:}
In addition to asking questions regarding participants' experience during the experiment, we asked participants if they could come up with real-life applications for this kind of remote collaborative MR system. Based on the responses we got from the participants, we identified three main types of applications where users foresee this setup being used. These include Collaborative Design and Prototyping (47.62\%), Remote Guidance and Education (14.29\%), and Entertainment (38.1\%) such as collaborative games. These categories align with the applications we identified to motivate this study.

We also asked participants to identify the possible issues that they expect in deploying a system of this kind in real-life scenarios. The answers revealed concerns regarding Unauthenticated Access (20\%), Network Reliability and Performance (45\%), and User Adoption (25\%). On the other hand, 10\% participants felt there were no major issues that should hinder the adoption of MR in real-life applications.

\section{Discussion}

\subsection{Low Recognition for Novel Attacks}
\emph{We identify that users are less likely to recognize attacks unique to MR and attribute them to other factors like technological glitches, partner miscommunication, and user errors.} This supports our hypothesis regarding \textbf{RQ2}. We believe the lack of recognition is due to MR environments typically being more cognitively demanding than 2D interfaces. The higher cognitive load might impair users' ability to recognize attacks overall. We also believe the misclassification of attack types by participants is due to a lack of familiarity with MR as they attributed most attacks to 2D concepts.

Specifically, the recognition frequency for the spatial occlusion attack was lower than that for the click redirection attack. This may stem from users' lack of prior understanding of the difference between a security threat and a technical glitch in immersive environments. Additionally, no participant hinted at the presence of a transparent object in the environment, suggesting that such concepts are difficult to grasp for users. Since the spatial occlusion attack relied heavily on MR-specific properties, its novelty and subtlety likely contributed to the users' responses and reduced recognition rates.

We observed that participants attributed the issues encountered during task completion to technological glitches and user errors, rather than suspecting malicious attacks. This observation provides key insights into user behavior and perception in collaborative MR environments and indicates that users might have a false sense of trust that these systems are more secure by design. This misplaced trust increases their vulnerability to MR-specific attacks. Additionally, users' tendency to attribute issues to technological glitches likely stems from their experience with conventional 2D computing, where technical problems are more common than UI-based security threats. As MR devices gain wider adoption, this prior experience may lead users to expect security measures in MR systems at par with those in traditional 2D environments. This expectation can potentially leave them vulnerable to MR-specific attacks.

\subsection{Discrepancy in Reported and Performed Mitigations} \,\,\,\,\,\,\emph{We~identified a gap between what users believe they did to mitigate certain attacks and their actions when subjected to attacks in a remote collaborative MR environment.} The discrepancy in reported and actual mitigation techniques of participants during the attacks revealed participants' lack of understanding of the environment and conceptualization of the 3D computing world. Higher recall of mitigation techniques in the click redirection condition implies that users have a stronger mental model of how to respond to this type of attack due to its similarity with the 2D computational environment and ability to visually observe the attack. In contrast, the greater discrepancy in reported techniques for mitigating spatial occlusion attacks suggests the attack may have been more complex and mentally demanding for the participants due to its subtlety. Additionally, the mitigation techniques used might have been a result of automatic response to the physical environment, rather than deliberate, conscious decisions. While intuitive mitigation techniques such as using near-interaction to grab the object may help resolve the immediate task, it can be more dangerous in practice because ignoring the subtle indicators of security threats means users might unknowingly continue interacting with malicious objects without realizing they are compromising security.

\subsection{Perceptions of Security Concerns}
Most participants mentioned remote guidance and design collaboration as potential applications of collaborative MR systems. Participants’ responses suggest a confident belief that remote collaborative MR systems are a viable candidate for these real-world applications in the near future. However, users mentioning network reliability and unauthenticated access as the two most common concerns during their interviews also highlight the increased challenges for the adoption of collaborative MR in real-world applications.
Participants' responses across the questionnaires also provide useful information with regard to security concerns, partially supporting the hypothesis for \textbf{RQ1}. Although only a few users identified the presence of the spatial occlusion attack, many still expressed concerns specifically about the security of the environment in the post-attack questionnaire after this condition. This suggests that, to some extent, users perceived the attack as spatial and immersive, even if they did not fully recognize its specific nature. On the other hand, click redirection attacks invoked users’ long-standing concerns about secure system access and data protection.

\subsection{Impact on Collaboration}
We also witnessed varying effects on collaboration in the presence of an attack. In the case of click redirection, as hypothesized, collaboration increased when the attack was encountered. This shows that being in a collaborative environment when faced with an attack helps discuss and mitigate issues. However, this is only true for situations when the presence of an attack has been recognized. On the contrary, when the attack had not been recognized it led to mistrust between the participants. We believe that the participants expected their partner to experience the same environment, and any discrepancies between their partner's description of the environment and their own experience of the environment led to confusion. This was most frequently observed during the spatial occlusion attack. This observation reinforced the notion that poor attack recognition can cause mistrust among collaborators in a shared environment.

\subsection{Self-Assessed Performance}
Participants underestimated the impact of the spatial occlusion attack, reporting slightly better results than the baseline condition. However, on average they took longer to complete tasks involving the spatial occlusion attack. This discrepancy in subjective performance rating and objective performance measurement suggests that the immersive nature of MR might have affected users' self-perception and awareness. The incorrect perceptions of individuals' performance can affect team dynamics in remote collaborative MR environments, as incorrect self-assessment can lead to misunderstandings or inefficiencies in collaborative tasks. 

Although the metrics for subjective performance ratings did not reveal significant differences across various attack conditions, emerging patterns suggest that these potential differences might become more evident with a larger sample size. Specifically, participants rated their performance as worst in the click redirection attack, where the attack was more noticeable. This suggests that while spatial occlusion attacks might not be as overtly noticeable as the click redirection attack, their subtlety can make them potentially more dangerous in practical applications.

\subsection{Design Recommendations} Our findings highlight the need to reflect on UI design in MR environments and educate users on potential MR-specific threats, building on the emphasis from \cite{rajaram2023eliciting} on integrating security and privacy considerations into collaborative MR systems while balancing usability, feasibility, and access control. We believe there is a pressing need to optimize UI design with intuitive security indicators that align with the user's mental model and cognitive load in MR space. We recommend that developers design applications to enhance visibility and understanding within their environment. This could include implementing controls and toggles that highlight all objects in the environment similar to basic 3D view management~\cite{bell2001view}, regardless of their opacity, and exposing transparent attack objects. Additionally, incorporating mandatory highlighting of the object being manipulated can help users identify which object they are interacting with. To further improve user awareness, spatial auditory cues could be used to indicate the location of the selected object in the 3D space. 
We also suggest that when a security vulnerability is suspected, developers go beyond traditional pop-up warnings or notifications and work on more immersive and integrated warning mechanisms that consider users’ cognitive load in MR space. There is also a clear need for user education and training in MR environments to help users recognize security risks. Training programs for critical MR applications such as remote medical operations should include a basic understanding of MR novelty, equipping users to recognize when something deviates from expected behavior and could indicate an MR security attack. Additionally, our findings support the work of Rajaram et al.~\cite{rajaram2023reframe}, who highlight the value of character-driven storyboarding in visualizing user interactions with security risks. This approach aids in developing user-centric solutions that improve awareness and understanding of privacy and security concerns in AR environments.

\subsection{Limitations and Future Work} While this study offers valuable insights into user behaviors in response to novel attacks within remote collaborative MR environments, it has some limitations. The sample size is limited due to the length of the user study and the need to recruit participants in pairs. Most of our results lacked statistical significance and demonstrated small effect sizes; thus, we report only the trending behaviors for those measures. We were limited by latency and errors in hand gesture detection with the Microsoft Hololens 2. The results might vary in terms of attack recognition if a newer device or a device with a controller is used. Additionally, participants lacked prior experience with head-worn MR devices, therefore studying behavior over longer periods of adaption may produce different results.

Future work should expand on the types of novel attacks that can be executed in remote collaborative MR environments. This will not only help us understand the threat landscape better but will also help gauge users' vulnerability to those attacks. Future studies should also investigate the factors that lead to discrepancies between perceived and actual performance. This could involve varying the types of tasks, the complexity of the environment, or the level of feedback provided to participants. We also suggest working towards effective solutions for these novel attacks by verifying the integrity of object manipulations through methods such as metamorphic testing~\cite{chen2018metamorphic, bose2025optimizing}. Researchers should investigate the effects of different training programs on users' ability to recognize and mitigate attacks. This could involve developing and testing various training modules and assessing their effectiveness through controlled experiments. It is also critical to make efforts towards improving security mechanisms within these environments and focus on designing user interfaces that make security issues more apparent and actionable.

\section{Conclusion}
Our research aimed to understand users' perspectives on security threats in remote collaborative MR environments. We developed novel MR-specific security attacks and tested them through a user study. While participants could identify certain attacks, we found that MR-specific threats were difficult for users to comprehend. Our recommendations offer design guidelines for developers to create systems that are both informative and resilient to such security threats. As MR technology continues to improve, its adoption across various real-world applications makes it essential to prioritize robust security measures and raise awareness among users.










\acknowledgments{
The authors acknowledge support from the Commonwealth Cyber Initiative and the Virginia Tech Institute for Creativity, Arts, and Technology. The authors also thank Dr. Maaz Gardezi
 and Dr. Andre Muelenaer for feedback on early prototypes and for motivating discussions on critical MR applications in the medical field. The authors also thank Sabaat Haroon for her contributions in conducting the pilot studies.}

\bibliographystyle{abbrv-doi}

\bibliography{template}

\begin{thebibliography}{10}

\bibitem{abraham2024you}
M.~Abraham, M.~Mcgill, and M.~Khamis.
\newblock What you experience is what we collect: User experience based fine-grained permissions for everyday augmented reality.
\newblock In {\em Proceedings of the CHI Conference on Human Factors in Computing Systems}, pp. 1--24, 2024.

\bibitem{abraham2022implications}
M.~Abraham, P.~Saeghe, M.~Mcgill, and M.~Khamis.
\newblock Implications of xr on privacy, security and behaviour: Insights from experts.
\newblock In {\em Nordic Human-Computer Interaction Conference}, pp. 1--12, 2022.

\bibitem{baker2018telemedicine}
J.~Baker and A.~Stanley.
\newblock Telemedicine technology: a review of services, equipment, and other aspects.
\newblock {\em Current allergy and asthma reports}, 18:1--8, 2018.

\bibitem{10.3389/feduc.2021.711619}
A.~Bashir, S.~Bashir, K.~Rana, P.~Lambert, and A.~Vernallis.
\newblock Post-covid-19 adaptations; the shifts towards online learning, hybrid course delivery and the implications for biosciences courses in the higher education setting.
\newblock {\em Frontiers in Education}, 6, 2021. doi: {{%
10\hspace{.1pt}\discretionary{.}{%
}{.}\hspace{.4pt}3389\discretionary{/}{%
}{/}feduc\hspace{.1pt}\discretionary{.}{%
}{.}\hspace{.4pt}2021\hspace{.1pt}\discretionary{.}{%
}{.}\hspace{.4pt}711619}}


\bibitem{bell2001view}
B.~Bell, S.~Feiner, and T.~H{\"o}llerer.
\newblock View management for virtual and augmented reality.
\newblock In {\em Proceedings of the 14th annual ACM symposium on User interface software and technology}, pp. 101--110, 2001.

\bibitem{bose2025optimizing}
D.~B. Bose, B.~David-John, and C.~Brown.
\newblock Optimizing ar application testing: Integrating metamorphic testing to address developer and end-user challenges.
\newblock In {\em International Conference on Human-Computer Interaction}, pp. 20--33. Springer, 2025.

\bibitem{article}
J.~Brooke.
\newblock Sus: A quick and dirty usability scale.
\newblock {\em Usability Eval. Ind.}, 189, 11 1995.

\bibitem{chen2024overload}
E.-C. Chen, P.-Y. Chen, I.~Chung, C.-R. Lee, et~al.
\newblock Overload: Latency attacks on object detection for edge devices.
\newblock In {\em Proceedings of the IEEE/CVF Conference on Computer Vision and Pattern Recognition}, pp. 24716--24725, 2024.

\bibitem{chen2018metamorphic}
T.~Y. Chen, F.-C. Kuo, H.~Liu, P.-L. Poon, D.~Towey, T.~Tse, and Z.~Q. Zhou.
\newblock Metamorphic testing: A review of challenges and opportunities.
\newblock {\em ACM Computing Surveys (CSUR)}, 51(1):1--27, 2018.

\bibitem{chenguser}
K.~Cheng, A.~Bhattacharya, M.~Lin, J.~Lee, A.~Kumar, J.~F. Tian, T.~Kohno, and F.~Roesner.
\newblock When the user is inside the user interface: An empirical study of ui security properties in augmented reality.

\bibitem{10.5555/3620237.3620289}
K.~Cheng, J.~F. Tian, T.~Kohno, and F.~Roesner.
\newblock Exploring user reactions and mental models towards perceptual manipulation attacks in mixed reality.
\newblock In {\em Proceedings of the 32nd USENIX Conference on Security Symposium}, SEC '23. USENIX Association, USA, 2023.

\bibitem{10.3934/ElectrEng.2019.2.181}
R.~A.~J. de~Belen, H.~Nguyen, D.~Filonik, D.~Del~Favero, and T.~Bednarz.
\newblock A systematic review of the current state of collaborative mixed reality technologies: 2013--2018.
\newblock {\em AIMS Electronics and Electrical Engineering}, 3(2):181--223, 2019.

\bibitem{10.1145/3359626}
J.~A. De~Guzman, K.~Thilakarathna, and A.~Seneviratne.
\newblock Security and privacy approaches in mixed reality: A literature survey.
\newblock {\em ACM Comput. Surv.}, 52(6), oct 2019. doi: {{%
10\hspace{.1pt}\discretionary{.}{%
}{.}\hspace{.4pt}1145\discretionary{/}{%
}{/}3359626}}


\bibitem{dell2022designing}
N.~Dell, D.~Estrin, H.~Haraldsson, and W.~Ju.
\newblock Designing extended reality guidance for physical caregiving tasks.
\newblock In {\em 2022 IEEE Conference on Virtual Reality and 3D User Interfaces Abstracts and Workshops (VRW)}, pp. 419--422. IEEE, 2022.

\bibitem{eom2024accuracy}
S.~Eom, T.~S. Ma, N.~Vutakuri, T.~Hu, A.~P. Haskell-Mendoza, D.~A. Sykes, M.~Gorlatova, and J.~Jackson.
\newblock Accuracy of routine external ventricular drain placement following a mixed reality--guided twist-drill craniostomy.
\newblock {\em Neurosurgical Focus}, 56(1):E11, 2024.

\bibitem{Erickson2020SharingGR}
A.~Erickson, N.~Norouzi, K.~Kim, R.~Schubert, J.~Jules, J.~J. Laviola, G.~Bruder, and G.~Welch.
\newblock Sharing gaze rays for visual target identification tasks in collaborative augmented reality.
\newblock {\em Journal on Multimodal User Interfaces}, 14:353 -- 371, 2020.

\bibitem{10049704}
C.~G. Fidalgo, Y.~Yan, H.~Cho, M.~Sousa, D.~Lindlbauer, and J.~Jorge.
\newblock A survey on remote assistance and training in mixed reality environments.
\newblock {\em IEEE Transactions on Visualization and Computer Graphics}, 29(5):2291--2303, 2023. doi: {{%
10\hspace{.1pt}\discretionary{.}{%
}{.}\hspace{.4pt}1109\discretionary{/}{%
}{/}TVCG\hspace{.1pt}\discretionary{.}{%
}{.}\hspace{.4pt}2023\hspace{.1pt}\discretionary{.}{%
}{.}\hspace{.4pt}3247081}}


\bibitem{gallardo2023speculative}
A.~Gallardo, C.~Choy, J.~Juneja, E.~Bozkir, C.~Cobb, L.~Bauer, and L.~Cranor.
\newblock Speculative privacy concerns about ar glasses data collection.
\newblock {\em Proceedings on Privacy Enhancing Technologies}, 2023.

\bibitem{gasques2021artemis}
D.~Gasques, J.~G. Johnson, T.~Sharkey, Y.~Feng, R.~Wang, Z.~R. Xu, E.~Zavala, Y.~Zhang, W.~Xie, X.~Zhang, et~al.
\newblock Artemis: A collaborative mixed-reality system for immersive surgical telementoring.
\newblock In {\em Proceedings of the 2021 CHI conference on human factors in computing systems}, pp. 1--14, 2021.

\bibitem{gasques2017exploring}
D.~Gasques~Rodrigues, A.~Jain, S.~R. Rick, L.~Shangley, P.~Suresh, and N.~Weibel.
\newblock Exploring mixed reality in specialized surgical environments.
\newblock In {\em Proceedings of the 2017 CHI Conference Extended Abstracts on Human Factors in Computing Systems}, pp. 2591--2598, 2017.

\bibitem{gugenheimer2022novel}
J.~Gugenheimer, W.-J. Tseng, A.~H. Mhaidli, J.~O. Rixen, M.~McGill, M.~Nebeling, M.~Khamis, F.~Schaub, and S.~Das.
\newblock Novel challenges of safety, security and privacy in extended reality.
\newblock In {\em CHI Conference on Human Factors in Computing Systems Extended Abstracts}, pp. 1--5, 2022.

\bibitem{10.3389/fict.2019.00005}
J.~Happa, M.~Glencross, and A.~Steed.
\newblock Cyber security threats and challenges in collaborative mixed-reality.
\newblock {\em Frontiers in ICT}, 6, 2019. doi: {{%
10\hspace{.1pt}\discretionary{.}{%
}{.}\hspace{.4pt}3389\discretionary{/}{%
}{/}fict\hspace{.1pt}\discretionary{.}{%
}{.}\hspace{.4pt}2019\hspace{.1pt}\discretionary{.}{%
}{.}\hspace{.4pt}00005}}


\bibitem{tlx}
S.~G. Hart.
\newblock Tlx.
\newblock Technical Report Report Number, NASA, 2000.
\newblock Available online.

\bibitem{10.1145/3631418}
X.~Huang and R.~Xiao.
\newblock Surfshare: Lightweight spatially consistent physical surface and virtual replica sharing with head-mounted mixed-reality.
\newblock {\em Proc. ACM Interact. Mob. Wearable Ubiquitous Technol.}, 7(4), jan 2024. doi: {{%
10\hspace{.1pt}\discretionary{.}{%
}{.}\hspace{.4pt}1145\discretionary{/}{%
}{/}3631418}}


\bibitem{johnson2018holocpr}
J.~G. Johnson, D.~G. Rodrigues, M.~Gubbala, and N.~Weibel.
\newblock Holocpr: Designing and evaluating a mixed reality interface for time-critical emergencies.
\newblock In {\em Proceedings of the 12th EAI International Conference on Pervasive Computing Technologies for Healthcare}, pp. 67--76, 2018.

\bibitem{mrc}
C.~Katins, P.~W. Woźniak, A.~Chen, I.~Tumay, L.~V.~T. Le, J.~Uschold, and T.~Kosch.
\newblock {Assessing User Apprehensions About Mixed Reality Artifacts and Applications: The Mixed Reality Concerns (MRC) Questionnaire}.
\newblock In {\em {Proceedings of the 2024 CHI Conference on Human Factors in Computing Systems}}, CHI '24. ACM, New York, NY, USA, 2024. doi: {{%
10\hspace{.1pt}\discretionary{.}{%
}{.}\hspace{.4pt}1145\discretionary{/}{%
}{/}3613904\hspace{.1pt}\discretionary{.}{%
}{.}\hspace{.4pt}3642631}}


\bibitem{kaufmann2003collaborative}
H.~Kaufmann.
\newblock Collaborative augmented reality in education.
\newblock {\em Institute of Software Technology and Interactive Systems, Vienna University of Technology}, pp. 2--4, 2003.

\bibitem{kilger2021detecting}
F.~Kilger, A.~Kabil, V.~Tippmann, G.~Klinker, and M.-O. Pahl.
\newblock Detecting and preventing faked mixed reality.
\newblock In {\em 2021 IEEE 4th International Conference on Multimedia Information Processing and Retrieval (MIPR)}, pp. 399--405. IEEE, 2021.

\bibitem{krauss2024makes}
V.~Krauss, P.~Saeghe, A.~Boden, M.~Khamis, M.~McGill, J.~Gugenheimer, and M.~Nebeling.
\newblock What makes xr dark? examining emerging dark patterns in augmented and virtual reality through expert co-design.
\newblock {\em ACM Transactions on Computer-Human Interaction}, 2024.

\bibitem{lebeck2018towards}
K.~Lebeck, K.~Ruth, T.~Kohno, and F.~Roesner.
\newblock Towards security and privacy for multi-user augmented reality: Foundations with end users.
\newblock In {\em 2018 IEEE Symposium on Security and Privacy (SP)}, pp. 392--408. IEEE, 2018.

\bibitem{masood2019augmented}
T.~Masood and J.~Egger.
\newblock Augmented reality in support of industry 4.0—implementation challenges and success factors.
\newblock {\em Robotics and Computer-Integrated Manufacturing}, 58:181--195, 2019.

\bibitem{mitrokotsa2007denial}
A.~Mitrokotsa and C.~Douligeris.
\newblock Denial-of-service attacks.
\newblock {\em Network Security: Current Status and Future Directions}, pp. 117--134, 2007.

\bibitem{mourtzis2021collaborative}
D.~Mourtzis, J.~Angelopoulos, and N.~Panopoulos.
\newblock Collaborative manufacturing design: a mixed reality and cloud-based framework for part design.
\newblock {\em Procedia CIRP}, 100:97--102, 2021.

\bibitem{o2023privacy}
J.~O'Hagan, P.~Saeghe, J.~Gugenheimer, D.~Medeiros, K.~Marky, M.~Khamis, and M.~McGill.
\newblock Privacy-enhancing technology and everyday augmented reality: Understanding bystanders' varying needs for awareness and consent.
\newblock {\em Proceedings of the ACM on Interactive, Mobile, Wearable and Ubiquitous Technologies}, 6(4):1--35, 2023.

\bibitem{paneva2024privacy}
V.~Paneva, M.~Strauss, V.~Winterhalter, S.~Schneegass, and F.~Alt.
\newblock Privacy in the metaverse.
\newblock {\em IEEE Pervasive Computing}, 23(3):73--78, 2024.

\bibitem{10.1145/3132787.3139200}
T.~Piumsomboon, A.~Day, B.~Ens, Y.~Lee, G.~Lee, and M.~Billinghurst.
\newblock Exploring enhancements for remote mixed reality collaboration.
\newblock In {\em SIGGRAPH Asia 2017 Mobile Graphics \& Interactive Applications}, SA '17. Association for Computing Machinery, New York, NY, USA, 2017. doi: {{%
10\hspace{.1pt}\discretionary{.}{%
}{.}\hspace{.4pt}1145\discretionary{/}{%
}{/}3132787\hspace{.1pt}\discretionary{.}{%
}{.}\hspace{.4pt}3139200}}


\bibitem{rajaram2023eliciting}
S.~Rajaram, C.~Chen, F.~Roesner, and M.~Nebeling.
\newblock Eliciting security \& privacy-informed sharing techniques for multi-user augmented reality.
\newblock In {\em Proceedings of the 2023 CHI Conference on Human Factors in Computing Systems}, pp. 1--17, 2023.

\bibitem{rajaram2023reframe}
S.~Rajaram, F.~Roesner, and M.~Nebeling.
\newblock Reframe: An augmented reality storyboarding tool for character-driven analysis of security \& privacy concerns.
\newblock In {\em Proceedings of the 36th Annual ACM Symposium on User Interface Software and Technology}, pp. 1--15, 2023.

\bibitem{rodrigues2018real}
D.~G. Rodrigues, J.~Johnson, and N.~Weibel.
\newblock Real-time guidance for cardiopulmonary resuscitation in mixed reality.
\newblock In {\em 12th EAI International Conference on Pervasive Computing Technologies for Healthcare--Demos, Posters, Doctoral Colloquium}, 2018.

\bibitem{ruth2019secure}
K.~Ruth, T.~Kohno, and F.~Roesner.
\newblock Secure $\{$Multi-User$\}$ content sharing for augmented reality applications.
\newblock In {\em 28th USENIX Security Symposium (USENIX Security 19)}, pp. 141--158, 2019.

\bibitem{saini2019you}
A.~Saini, M.~S. Gaur, V.~Laxmi, and M.~Conti.
\newblock You click, i steal: analyzing and detecting click hijacking attacks in web pages.
\newblock {\em International Journal of Information Security}, 18:481--504, 2019.

\bibitem{santos2007improve}
P.~Santos, A.~Stork, T.~Gierlinger, A.~Pagani, C.~Paloc, I.~Barandarian, G.~Conti, R.~de~Amicis, M.~Witzel, O.~Machui, et~al.
\newblock Improve: An innovative application for collaborative mobile mixed reality design review.
\newblock {\em International Journal on Interactive Design and Manufacturing (IJIDeM)}, 1:115--126, 2007.

\bibitem{slocum2024doesnt}
C.~Slocum, Y.~Zhang, E.~Shayegani, P.~Zaree, N.~Abu-Ghazaleh, and J.~Chen.
\newblock That doesn't go there: Attacks on shared state in multi-user augmented reality applications, 2024.

\bibitem{syal2020threats}
S.~Syal and R.~Mathew.
\newblock Threats faced by mixed reality and countermeasures.
\newblock {\em Procedia Computer Science}, 171:2720--2728, 2020.

\bibitem{9963687}
S.~Valluripally, B.~Frailey, B.~Kruse, B.~Palipatana, R.~Oruche, A.~Gulhane, K.~A. Hoque, and P.~Calyam.
\newblock Detection of security and privacy attacks disrupting user immersive experience in virtual reality learning environments.
\newblock {\em IEEE Transactions on Services Computing}, 16(4):2559--2574, 2023. doi: {{%
10\hspace{.1pt}\discretionary{.}{%
}{.}\hspace{.4pt}1109\discretionary{/}{%
}{/}TSC\hspace{.1pt}\discretionary{.}{%
}{.}\hspace{.4pt}2022\hspace{.1pt}\discretionary{.}{%
}{.}\hspace{.4pt}3216539}}


\bibitem{vrchat}
{VRChat Inc.}
\newblock Vrchat: A social virtual reality platform, 2017.
\newblock A platform for social interaction and user-generated content in virtual reality.

\bibitem{opensoundcontrol}
M.~Wright and A.~Freed.
\newblock Open sound control (osc).
\newblock Protocol, 1997.
\newblock A protocol for communication among computers, sound synthesizers, and other multimedia devices. Available at \url{https://opensoundcontrol.org/}.

\bibitem{yang2024can}
W.~Yang, X.~Dengxiong, X.~Wang, Y.~Hu, and Y.~Zhang.
\newblock “i can see your password”: A case study about cybersecurity risks in mid-air interactions of mixed reality-based smart manufacturing applications.
\newblock {\em Journal of Computing and Information Science in Engineering}, 24(3):031004, 2024.

\bibitem{zhang2023s}
Y.~Zhang, C.~Slocum, J.~Chen, and N.~Abu-Ghazaleh.
\newblock It's all in your head (set): Side-channel attacks on $\{$AR/VR$\}$ systems.
\newblock In {\em 32nd USENIX Security Symposium (USENIX Security 23)}, pp. 3979--3996, 2023.

\end{thebibliography}
\end{document}